%% file: text.tex
\documentclass[12pt,14paper]{article}
\usepackage[english]{babel}
\usepackage[latin1]{inputenc}
\usepackage[normalem]{ulem}
\usepackage[intlimits]{amsmath}
\usepackage{amssymb}
\usepackage{placeins}
\usepackage{graphics}
\usepackage{graphicx}
\usepackage{geometry}
\usepackage{color}
\usepackage{axodraw}
\usepackage{verbatim} 
\usepackage{multirow} 
\usepackage{soul}

\newcommand{\GeV}{\,\mbox{GeV}}
\newcommand{\TeV}{\,\mbox{TeV}}

\newcommand{\cm}{\,\mbox{cm}}
\newcommand{\s}{\,\mbox{s}}

\begin{document}

\title{\vspace{-2cm} 
{\normalsize
\flushright TUM-HEP 894/13\\}
\vspace{0.6cm} 
\bf 
Novel Gamma-ray Spectral Features \\ 
in the Inert Doublet Model\\[8mm]}

\author{Camilo Garcia-Cely and Alejandro Ibarra \\[2mm]
{\normalsize\it Physik-Department T30d, Technische Universit\"at M\"unchen,}\\[-0.05cm]
{\it\normalsize James-Franck-Stra\ss{}e, 85748 Garching, Germany}
}

\maketitle

\begin{abstract}
The inert doublet model contains a neutral stable particle which is an excellent dark matter candidate. We discuss in this paper the indirect signatures of this model in gamma-rays when the dark matter mass is larger than the $W$ boson mass. We show that, in addition to the featureless gamma-ray spectrum produced in the annihilations into two weak gauge bosons, the model generically predicts a distinctive spectral feature from the internal bremsstrahlung process $H^0 H^0 \rightarrow W^+ W^- \gamma$. We discuss under which conditions the spectral feature is generated and we construct a number of benchmark points, compatible with the observed relic density and all other direct and indirect detection experiments, which lead to a sharp gamma-ray feature from internal bremsstrahlung. 
\end{abstract}

\section{Introduction}
\label{sec:introduction}

The inert doublet model \cite{Deshpande:1977rw,Ma:2006km,Barbieri:2006dq} is a minimal extension of the Standard Model which accounts for the dark matter of the Universe and which consists in introducing one extra scalar field, odd under an unbroken $Z_2$ symmetry, with identical gauge quantum number as the Standard Model Higgs. The lightest among the particles of the new sector is then stable on cosmological scales and, if electrically neutral, constitutes an excellent dark matter candidate. Indeed, it has been shown that this model can reproduce the correct dark matter relic density in a low mass regime, $10 \GeV\lesssim M_{\rm DM}\lesssim M_W$, and a high mass regime, $M_{\rm DM}\gtrsim 500\GeV$, for reasonable values of the model parameters and without fine-tunings, as well as in an intermediate mass regime, $M_W \lesssim M_{\rm DM}\lesssim 150 \GeV$, for special choices of the model parameters. However, as for all other dark matter models proposed, no signature of the inert Higgs doublet model has been observed apart from the gravitational effects of the dark matter particles in the dynamics of galaxies, clusters of galaxies or the Universe at large scale. The inert Higgs doublet model offers, however, a rich phenomenology and predicts potentially observable effects in direct and indirect dark matter searches \cite{Barbieri:2006dq,LopezHonorez:2006gr,Majumdar:2006nt,Gustafsson:2007pc,Agrawal:2008xz,Andreas:2009hj,Nezri:2009jd,Arina:2009um,Honorez:2010re,LopezHonorez:2010tb,Klasen:2013btp}, collider searches \cite{Barbieri:2006dq,Cao:2007rm,Lundstrom:2008ai,Dolle:2009ft,Miao:2010rg}, electroweak precision tests \cite{Barbieri:2006dq,Grimus:2008nb} or the Higgs diphoton decay rate \cite{Arhrib:2012ia,Swiezewska:2012eh,Goudelis:2013uca}.

In this work we will focus in the possibility of detecting signatures of the inert Higgs doublet model through the observation of the gamma-rays which are predicted to be produced in the dark matter annihilations. More specifically, in the high mass regime, the dark matter particle annihilates mostly in the channels $H^0 H^0\rightarrow W^+ W^-, Z^0 Z^0, h h, t \bar t$ while in the low mass regime, when the previous decay channels are kinematically forbidden, mostly through $H^0 H^0 \rightarrow b \bar b$ (three body annihilations of the type $H^0 H^0\rightarrow W W^*\rightarrow W f \bar f'$ can also be important in some regions of the parameter space \cite{Honorez:2010re}). The hadronization of the weak gauge bosons, Higgs boson or quarks produces neutral pions, which in turn decay into photons generating a gamma-ray flux with a featureless spectrum which could be detected at the Earth. For reasonable parameters, the gamma-ray flux from dark matter annihilations is significantly smaller than the, also featureless, diffuse galactic gamma-ray emission. Hence disentangling a dark matter signal from the astrophysical background requires a very good understanding of the latter, which is unfortunately still lacking.

A promising strategy to indirectly search for dark matter annihilations is then the search for spectral features, which might stand out over the featureless diffuse background even for moderate values of the annihilation cross section~\cite{Srednicki:1985sf,Rudaz:1986db,Bergstrom:1988fp,Bergstrom:1989jr,Flores:1989ru,Bringmann:2007nk,Ibarra:2012dw}. Furthermore, the observation of gamma-ray features in the energy spectrum would constitute a strong hint for an exotic contribution in the gamma-ray flux. It has been pointed out that the inert doublet model produces gamma-ray lines in the annihilations $H^0 H^0\rightarrow \gamma \gamma, \gamma Z$~\cite{Gustafsson:2007pc}. These processes arise at the one loop level and therefore the expected cross section is rather suppressed, $\sigma v = {\cal O}(10^{-28}) \cm^3 \s^{-1}$ in the low mass regime~\cite{Gustafsson:2007pc}, which is a factor of a few below the sensitivity presently attained by the Fermi-LAT~\cite{Ackermann:2012qk}.

In this paper we will focus on another class of spectral features, stemming from the three body annihilation $H^0 H^0\rightarrow W^+W^- \gamma$. This process is generated through the exchange of a charged scalar in the t-channel and therefore produces, as discussed in \cite{Bringmann:2007nk}, a bump close to the kinematical endpoint of the gamma-ray spectrum when the mass of the particle in the t-channel is degenerated to the dark matter mass. This situation naturally occurs in the high mass regime of the inert doublet model, therefore this signature is generically expected to appear. We will review in Section \ref{sec:IDM} the main features of the inert doublet dark matter model. Then, in Section \ref{sec:IB} we will analyze the gamma-ray production via the internal bremsstrahlung and in Section \ref{sec:gamma-signals} the possibility of detecting the associated spectral feature in gamma-ray telescopes. Lastly, we will present our conclusions in Section \ref{sec:conclusions} and  an Appendix containing details of the calculation of the annihilation cross section in the internal bremsstrahlung process.

\section{The inert doublet Model}
\label{sec:IDM}

We consider the Inert Doublet Model (IDM), where the particle content of the Standard Model is extended by one complex scalar field $\eta$, which is a singlet under $SU(3)_C$, doublet under $SU(2)_L$ and has hypercharge $1/2$. Furthermore, we postulate a discrete  $Z_2$ symmetry under which the Standard Model particles are even while the extra scalar $\eta$ is odd. With this particle content the Lagrangian can be cast as ${\cal L}={\cal L}_{\rm SM}+{\cal L}_\eta$, where ${\cal L}_{\rm SM}$ is the Standard Model Lagrangian including a potential for the Higgs doublet $\Phi$
\begin{equation}
{\cal L}_{\rm SM} \supset - m_1^2 \Phi^\dagger \Phi - \lambda_1 (\Phi^\dagger \Phi)^2 \;,
\end{equation}
and ${\cal L}_\eta$ is the most general $Z_2$ invariant Lagrangian involving the scalar doublet $\eta$ 
\begin{eqnarray}
{\cal L}_\eta&=&(D_\mu \eta)^\dagger  (D^\mu \eta) -m_2^2\eta^\dagger \eta  
- \lambda_2(\eta^\dagger \eta)^2
- \lambda_3(\Phi^\dagger \Phi)(\eta^\dagger \eta) \nonumber\\
&& - \lambda_4(\Phi^\dagger \eta)(\eta^\dagger \Phi) 
-\dfrac{1}{2} \left( \lambda_5(\Phi^\dagger \eta)(\Phi^\dagger \eta)+{\rm \text{\small h.c.}}\right) \;,
\end{eqnarray}
where $D_\mu$ denotes the covariant derivative. Notice that all mass parameters and quartic couplings can be taken real and that CP is thus conserved.
After the electroweak symmetry breaking, and assuming that the $Z_2$ symmetry remains unbroken, only the Higgs doublet acquires an expectation value,
\begin{equation}
\langle \Phi \rangle =\begin{pmatrix} 0 \\ \langle \Phi^0\rangle \end{pmatrix} \;, \hspace{40pt} \langle \eta \rangle = \begin{pmatrix} 0 \\ 0 \end{pmatrix}  \;,
\end{equation}
where $\langle \Phi^0\rangle =\sqrt{-\frac{m_1^2}{2\lambda_1}} \approx 174\GeV$. Besides the standard model Higgs particle $h$, the other scalar particles  from the extra doublet $\eta= \begin{pmatrix} H^+ \\ \frac{1}{\sqrt2} \left( H^0 + i A^0 \right) \end{pmatrix}$ are two charged states $H^\pm$, one CP-even neutral state $H^0$ and one CP-odd neutral state $A^0$. In terms of these fields, the scalar potential in the unitary gauge reads
\begin{eqnarray}
V_{\rm Scalar} &=& \frac{1}{2}\left(M_h^2 h^2 + M_H^{0} {H^0}^2+ M_{A^0}^2 {A^0}^2 + M_{H^+}^2 H^+H^-\right) 
+\lambda_1 \left(\frac{1}{4} h^4 + \sqrt{2}\langle \Phi^0\rangle  h^3 \right) \nonumber\\
&+& \lambda_2  \left(\frac{1}{2}{A^0}^2+\frac{1}{2} {H^0}^2+ H^+ H^-\right)^2+ \left(\frac{1}{2}h^2+ \sqrt{2} \langle \Phi^0\rangle h \right)\left( \lambda_{A^0}  {A^0}^2+ \lambda_{H^0} {H^0}^2+ \lambda_3  H^+ H^- \right)   \;,\nonumber
\label{scalarpotential}
\end{eqnarray}
where
\begin{eqnarray}
M_h^2 = -2 m_1^2 \;, \hspace{15pt} 
M^2_{H^0} &=& m_2^2 + 2 \lambda_{H^0} \langle \Phi^0\rangle^2 \;,\hspace{15pt}
M^2_{A^0} = m_2^2 +2  \lambda_{A^0} \langle \Phi^0\rangle^2\;, \hspace{15pt}
M^2_{H^+} = m_2^2 + \lambda_3 \langle \Phi^0\rangle^2\;, \nonumber\\
\lambda_{H^0} &=& \frac{1}{2} (\lambda_3+\lambda_4+\lambda_5) \;, \hspace{15pt} 
\lambda_{A^0} = \frac{1}{2} (\lambda_3+\lambda_4-\lambda_5)\;.
\label{masses}
\end{eqnarray}
Due to the existence of the $Z_2$ symmetry, the lightest among the extra particles is stable and, if neutral, also a dark matter candidate. In fact, the role of $H^0$ and $A^0$ can be interchanged by performing $\left(H^0 , A^0\right) \to \left(-A^0, H^0\right)$ and $\lambda_5 \to -\lambda_5$. We will assume in what follows, and without loss of generality, that $H^0$ is the dark matter candidate, which occurs, according to eq.(\ref{masses}), when $\lambda_5<0$. 

The scalar potential eq.(\ref{scalarpotential}) is determined by seven independent parameters. Two of them are already known: they are the Higgs boson mass --  which will be assumed equal to $M_h \approx 125 \GeV$ -- and the vacuum expectation value of the Higgs field. These two are related to the Lagrangian parameters $m_1$ and $\lambda_1$. Among the remaining five parameters, only one is dimensionful and can be chosen to be the dark matter mass $M_{H^0}$. The other independent parameters can be chosen as $\lambda_2$,  $\lambda_3, \lambda_4$ and $\lambda_5$. Notice that the last two determine the mass splittings between the different exotic particles, concretely
\begin{equation}
M^2_{H^+} = M^2_{H^0} - (\lambda_4+\lambda_5)\langle \Phi^0\rangle^2 \;, \hspace{15pt}
M^2_{A^0} = M^2_{H^0} - 2 \lambda_5 \langle \Phi^0\rangle^2\;.
\label{masssplitting}
\end{equation}

The quartic couplings are constrained from the requirement of vacuum stability \cite{Gunion:2002zf}
\begin{equation}
\lambda_1 > 0\;, \hspace{20pt}\lambda_2 > 0\;, \hspace{20pt}\lambda_3 > -2 (\lambda_1\lambda_2)^{\frac{1}{2}}\;, \hspace{20pt}\lambda_3+\lambda_4-|\lambda_5| > -2 (\lambda_1\lambda_2)^{\frac{1}{2}}\;.
\label{treelevelconstraint}
\end{equation}
Besides, the unitarity of the $S$-matrix for scalar-to-scalar scattering sets upper limits on certain combinations of couplings~\cite{Ginzburg:2004vp,Branco:2011iw}
\begin{eqnarray}
\lambda_1+\lambda_2 \pm \sqrt{(\lambda_1-\lambda_2)^2+\lambda_4^2} < 8\pi, && \lambda_3 \pm \lambda_4 < 8 \pi \nonumber  \\
\lambda_1+\lambda_2 \pm \sqrt{(\lambda_1-\lambda_2)^2+\lambda_5^2} < 8\pi, && \lambda_3 \pm \lambda_5 < 8 \pi   \\
3(\lambda_1+\lambda_2) \pm \sqrt{9(\lambda_1-\lambda_2)^2+(2\lambda_3+\lambda_4)^2} < 8\pi, && 
\lambda_3+2\lambda_4 \pm 3\lambda_5 < 8 \pi\;.  \nonumber
\end{eqnarray}
Lastly, it is common to impose also  perturbativity on the parameters of the model. This condition along with the vacuum stability requirement significantly  constrain the mass splittings among the exotic particles. Moreover, for a heavy exotic neutral Higgs ($M_{H^0} \gg M_W$) the splitting is relatively small and we expect the particles belonging to the extra doublet to have nearly degenerate masses. This is consistent with the fact that at very high energies electroweak symmetry breaking effects are negligible and that therefore the members of any $SU(2)\times U(1)$ multiplet should have similar masses.

Many works have studied the relic abundance of $H^0$ \cite{Barbieri:2006dq,LopezHonorez:2006gr,LopezHonorez:2010tb,Cirelli:2005uq,Hambye:2007vf,Hambye:2009pw} and have shown that the measured cold dark matter abundance, $\Omega h^2=0.1199\pm 0.0027$ \cite{Ade:2013lta} could be reproduced in three scenarios: a low mass regime $ 10\GeV \lesssim M_{H^0} \lesssim M_W$, a high mass regime $M_{H^0} \gtrsim 500\GeV$, and an intermediate mass regime $M_W \lesssim M_{H^0}\lesssim 150 \GeV$. In the low mass regime the annihilation channels into weak gauge bosons or Higgs bosons are kinematically closed and only the annihilations into light fermions, with a rate controlled by the size of the quartic couplings, are possible. The correct relic density can then be generated by adjusting the values of the quartic couplings, possibly leading to large mass splittings among the scalar mass eigenstates. For dark matter masses above the $W$ and $Z$ mass thresholds, the annihilations into weak gauge bosons can be very efficient in general and in fact generate a dark matter density below the measured value for $M_{H^0}\lesssim 500  \GeV$. Nevertheless, under certain circumstances, in the region $M_W \lesssim M_{H^0}\lesssim 150 \GeV$, some contributions from the annihilation diagrams cancel each other out, allowing for another viable region where it is possible to reproduce the observed relic density \cite{LopezHonorez:2010tb}. Lastly, when $M_{H^0}\gtrsim 500  \GeV$, following the unitarity requirement that the cross section must decrease with the inverse of $M_{H^0}^{2}$, the annihilation rate is again small allowing to reproduce the correct dark matter abundance.. In this case, since $M_{H^0}\gg \langle \Phi^0\rangle$, all exotic scalar states have fairly degenerate masses. In this paper we will focus in the latter case, where the dark matter particle annihilates into weak gauge bosons via the t-channel exchange of a charged scalar which is fairly degenerate in mass with the dark matter particle. Under these conditions, it is expected an enhancement of the internal bremsstrahlung in the annihilation, which in turn might lead to an intense feature in the gamma-ray spectrum.

\section{Internal bremsstrahlung in the inert doublet model}
\label{sec:IB}

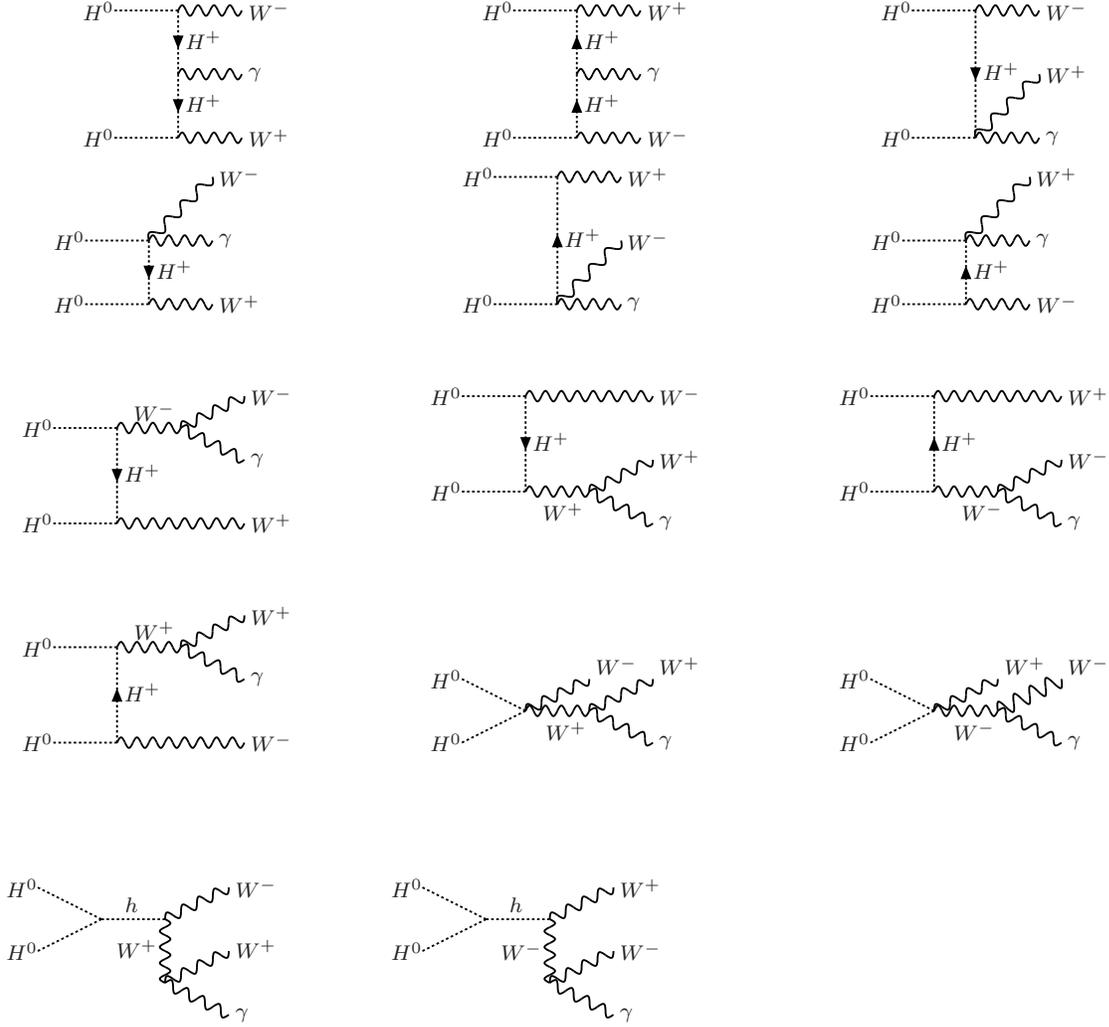
\begin{figure}[t]
\input{figure1.tex}
\caption{\footnotesize Feynman diagrams contributing to the annihilation process $H^0 H^0\rightarrow W^+ W^- \gamma$ in the unitary gauge.}
\label{fig:diagramswwg}
\end{figure}

The annihilation process into $W$ bosons with the associated emission of a photon, $H^0 H^0\rightarrow W^+ W^- \gamma$, is described in the unitary gauge by the fourteen diagrams shown in Fig. \ref{fig:diagramswwg}. The the first twelve diagrams are proportional to $g^2 e$ while the last two are proportional to $g \lambda_H e$, where $g$ and $e$ are, respectively, the weak and the electromagnetic coupling constants and $\lambda_H$ is the quartic coupling entering in the $H^0 H^0 h$ vertex. Note that, following Eq.(\ref{masses}),
\begin{equation}
\lambda_{H^0}=\frac{1}{2}\left(\lambda_3+\frac{M_{H^0}^2-M_{H^+}^2}{\langle \Phi^0\rangle^2}\right)\;.
\end{equation}
Then, the amplitudes of the last two diagrams depend on the quartic coupling $\lambda_3$ and on the mass splitting between the exotic charged and neutral scalars. 

As discussed in detail in the Appendix, the differential velocity weighted annihilation cross section can be cast as:
\begin{equation}
\frac{d(\sigma v)_{W^+ W^- \gamma}}{dx}  = 
\frac{d(\sigma v)}{dx}\Bigg|_{\text{Gauge}}+
\frac{d(\sigma v)}{dx}\Bigg|_{\text{Quartic}}+
\frac{d(\sigma v)}{dx}\Bigg|_{\text{Interference}}\;,
\label{eq:crosssection}
\end{equation}
where each term is separately gauge invariant. In the unitary gauge, the part labelled as ``gauge'' receives contributions from the diagrams with a charged scalar in the t-channel and generates, in addition to the usual contribution from final state radiation, a spectral feature which is sharper when $M_{H^0}=M_{H^+}$, since in this regime the scalar propagator is enhanced for large photon energies~\cite{Bringmann:2007nk}. The piece labelled as ``quartic'', on the other hand, receives contributions from the diagrams with the light Higgs in the s-channel and leads to a spectrum without distinctive spectral features. Then, the shape of the differential photon spectrum from internal bremsstrahlung essentially depends on the relative weight of the gauge and the quartic contributions to the cross section, which is in turn determined by the quartic coupling $\lambda_3$ and by the mass splitting between $H^0$ and $H^+$ (the contribution labelled as ``interference'' can be safely neglected, as shown in the Appendix).

We show in Figure \ref{fig:lambda3dependence}, upper plots, the three different contributions in Eq.~(\ref{eq:crosssection}) as a function of $x=E_\gamma/M_{H^0}$ in the limit $M_{H^0}=M_{H^+}$, which is a good approximation in the high mass regime of the inert Higgs dark matter model, for the cases $M_{H^0}$=0.5 TeV, 1 TeV and 5 TeV and for different values of the quartic coupling $\lambda_3$. The analytic result for the differential annihilation cross section in this limit is rather complicated and is presented in the Appendix.\footnote{We have used CalcHEP \cite{Pukhov:1999gg, Pukhov:2004ca} for parts of the analytical as well as for numerical computations.} We also multiply the spectrum by $x^2$ to emphasize the spectral structure. The blue, green and red lines represent, respectively, the gauge, quartic and interference terms, the latter in absolute value. Besides, the darkest lines correspond to $|\lambda_3|=0$ and the lines become lighter as $|\lambda_3|$ is increased in intervals of 0.4, the lightest lines corresponding to $|\lambda_3|=2$. The pure gauge part produces a spectrum that depends only on the dark matter mass and that displays a feature close to the endpoint of the spectrum  which becomes sharper and sharper as $M_{H^0}$ increases. The quartic part is proportional to $\lambda_3^2$ and becomes more and more important as $|\lambda_3|$ increases, eventually dominating over the gauge part for values of $x$ closer and closer to one. For a dark matter mass $M_{H^0}=0.5$ TeV and $|\lambda_3|=2$ the sharp spectral feature is practically erased in the total spectrum due to the effect of the final state radiation, however, for large dark matter masses the sharp spectral feature remains clearly visible even for $|\lambda_3|=2$. This behaviour can be better appreciated in Fig.\ref{fig:lambda3dependence}, lower plots, where we show the photon multiplicity from internal bremsstrahlung, defined as:
\begin{equation}
  \frac{dN^{\rm IB}_{\rm W^+ W^- \gamma}}{dx} = \frac{1}{(\sigma v)_{W^+W^-}}
\frac{d(\sigma v)_{W^+W^-\gamma} }{dx}\;,
\end{equation}
where $d(\sigma v)_{W^+W^-\gamma}/dx$ is given in Eq.~(\ref{eq:crosssection}). As before, the darkest line corresponds to $|\lambda_3|=0$ and the lightest to $|\lambda_3|=2$ and the intermediate lines correspond to changing $\lambda_3$ in intervals of 0.4.

\begin{figure}
\begin{center}
\includegraphics[width=0.32\textwidth]{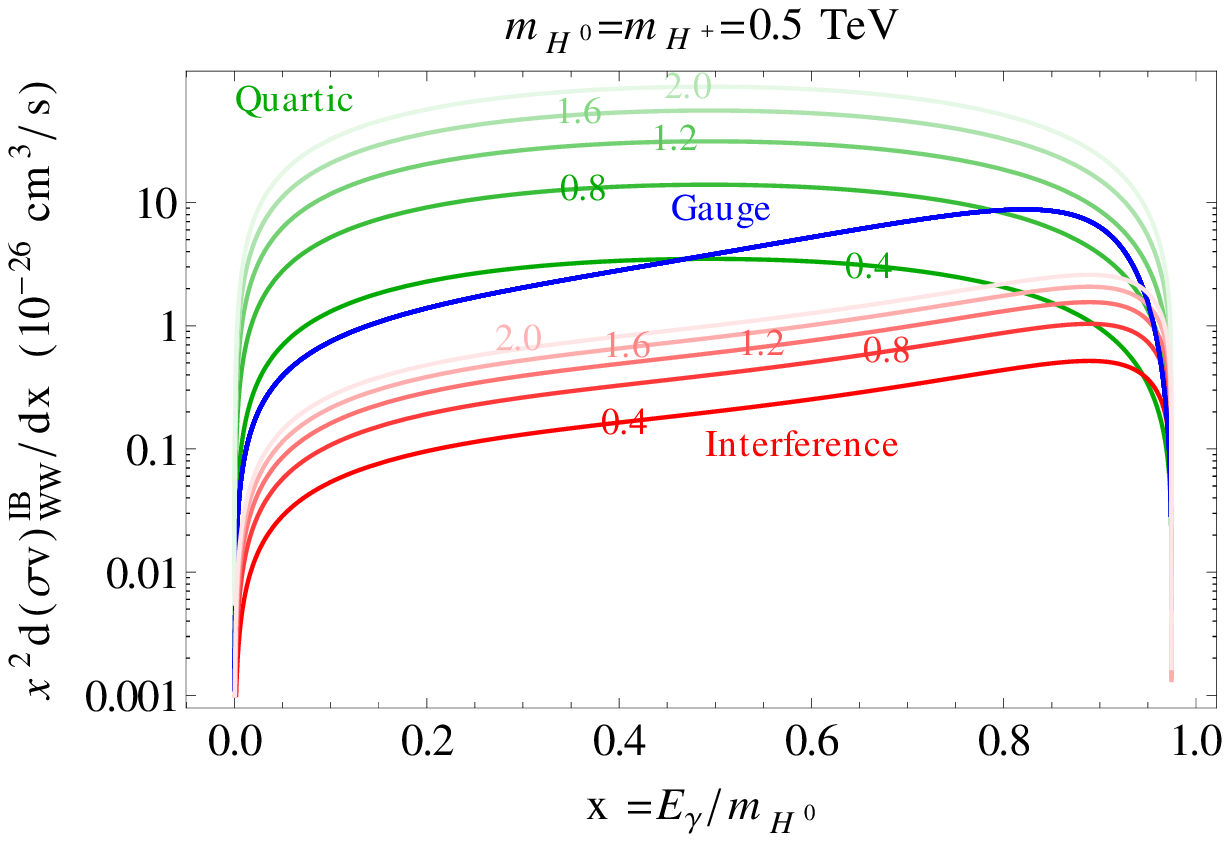}
\includegraphics[width=0.32\textwidth]{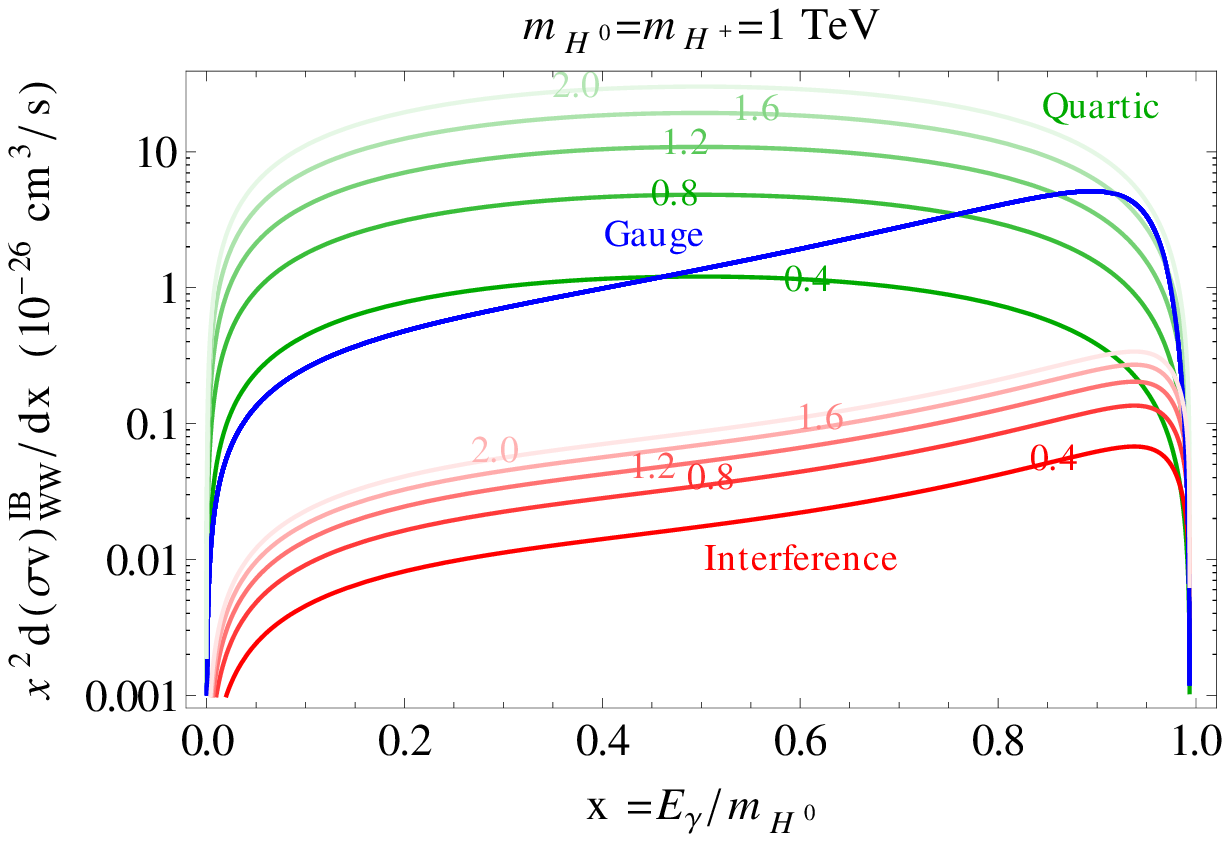}
\includegraphics[width=0.32\textwidth]{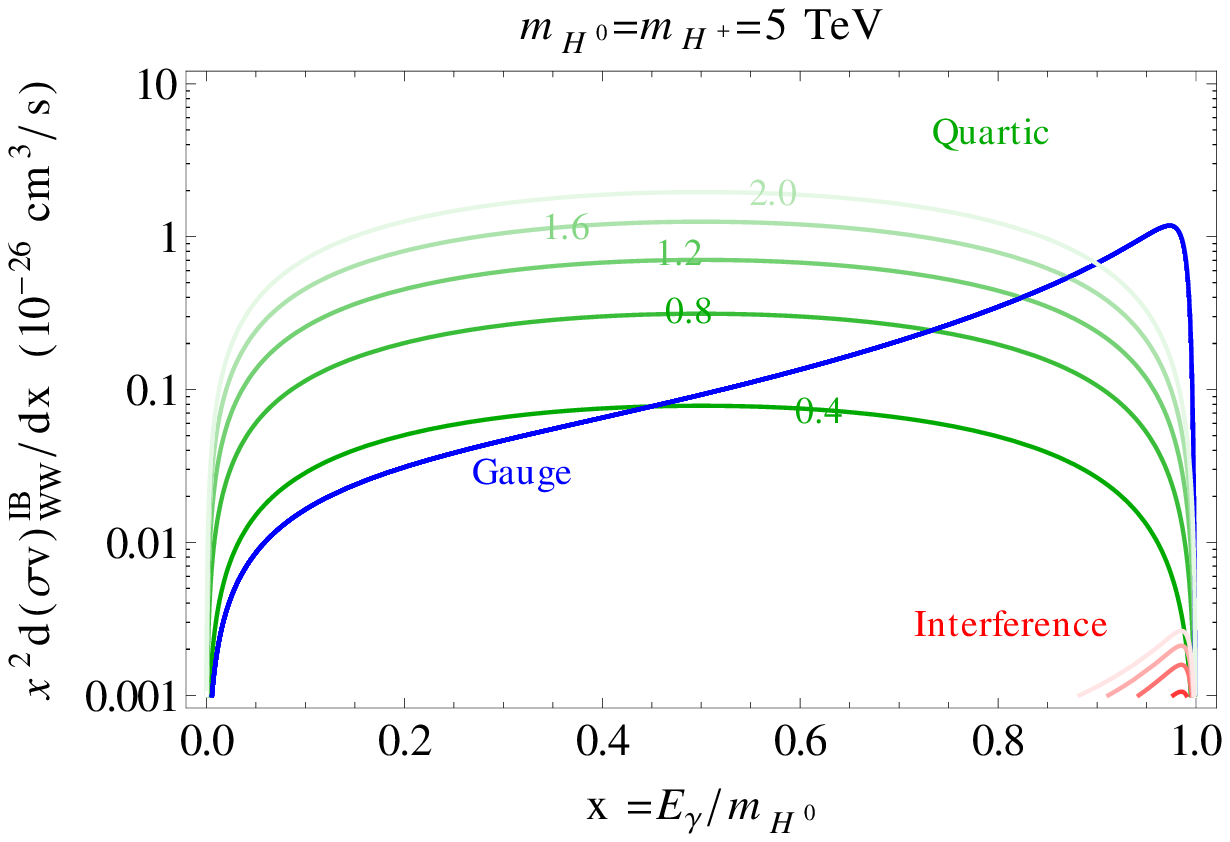} \\
\includegraphics[width=0.32\textwidth]{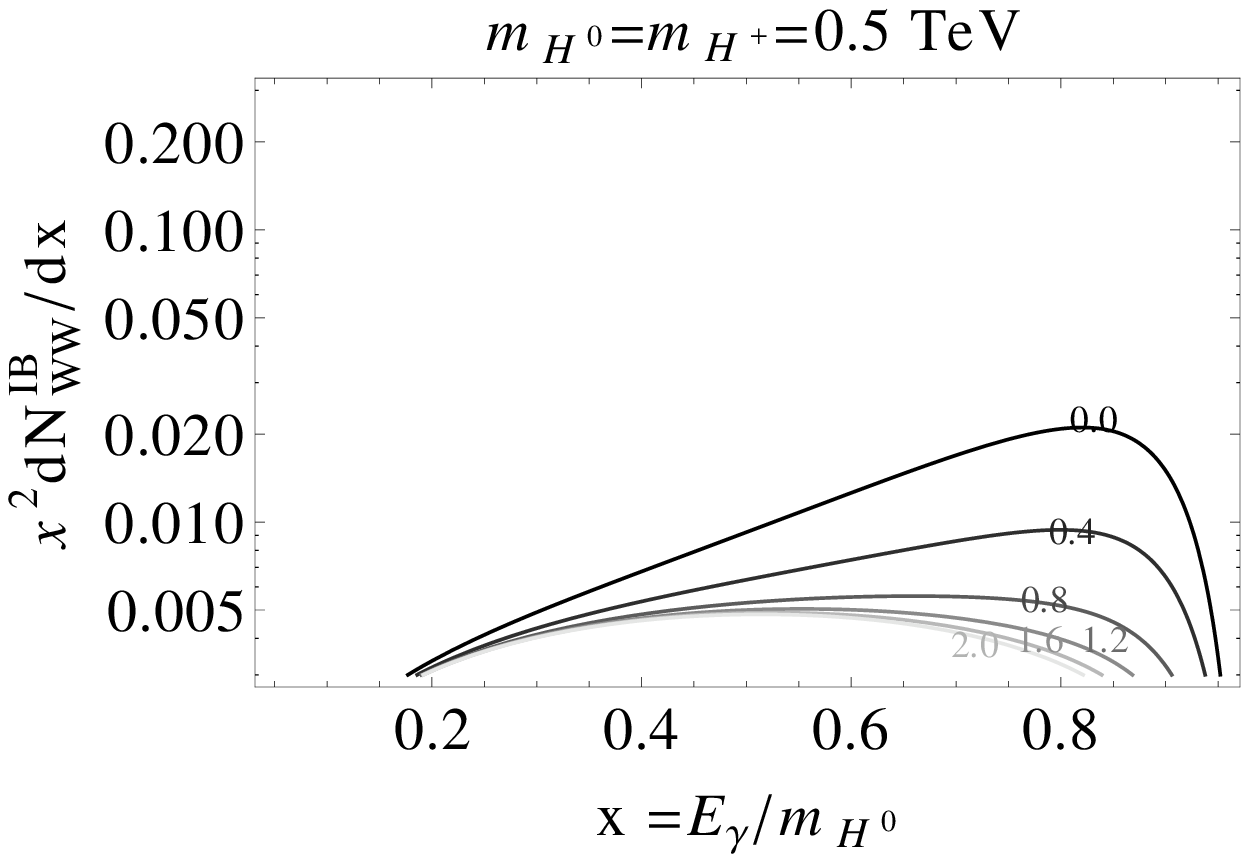}
\includegraphics[width=0.32\textwidth]{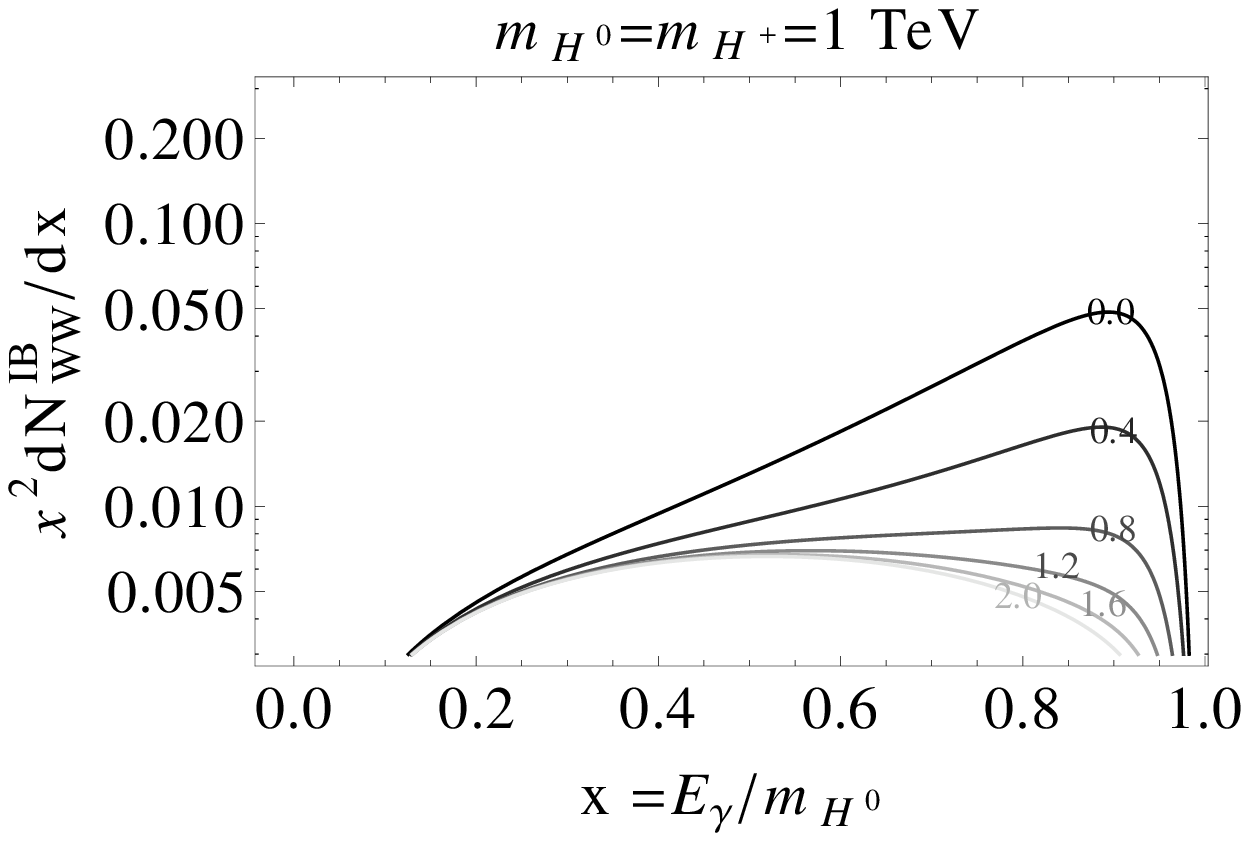}
\includegraphics[width=0.32\textwidth]{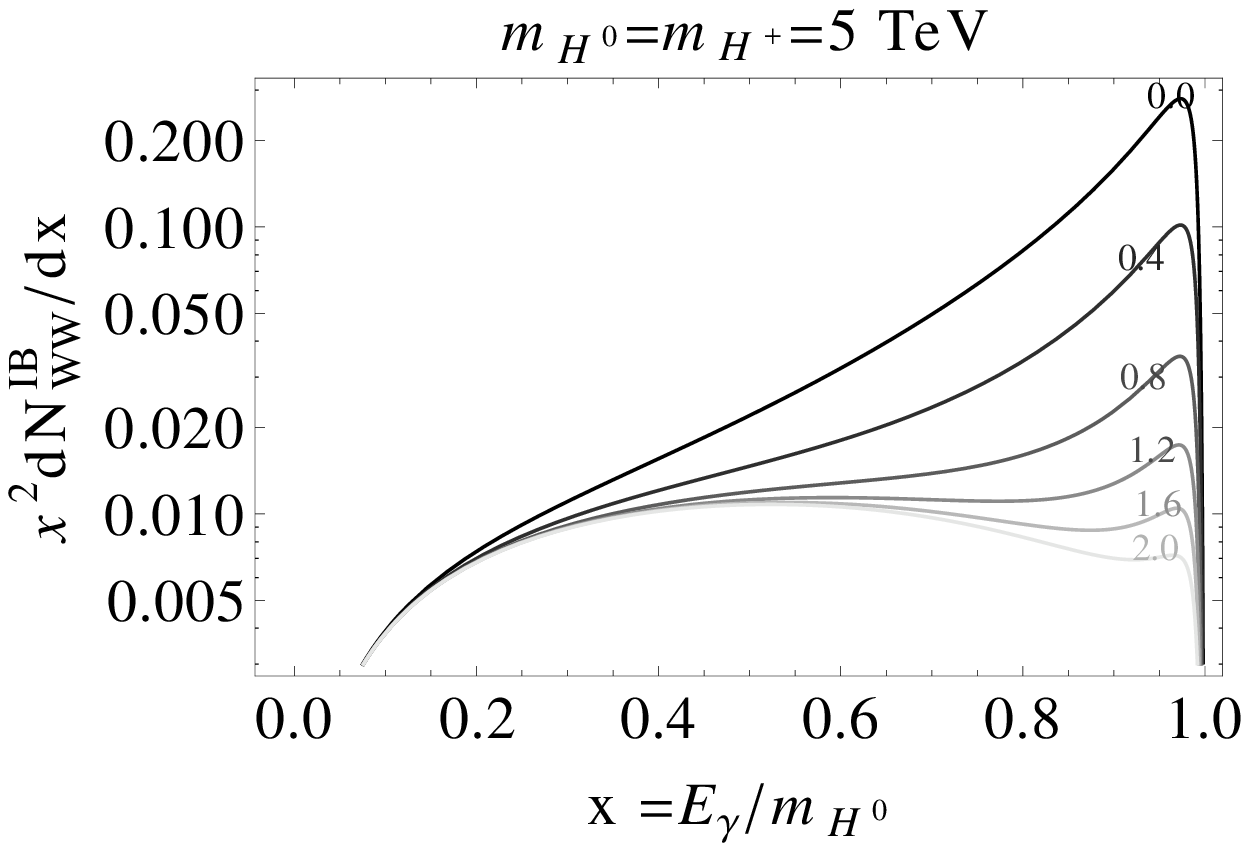}
\caption{\footnotesize Contributions to the cross section (top plots) and the total multiplicity of the hard photon (bottom plots) corresponding to the internal bremsstrahlung process $H^0 H^0\rightarrow W^+W^-\gamma$, when the charged scalar is degenerate in mass with the dark matter particle and $m_{H^0}=0.5$ TeV (left plots), 1 TeV (middle plots) and 5 TeV (right plots). In the top plot, the blue, green and red lines correspond, respectively, to the gauge, quartic and interference terms, the latter in absolute value. The different darknesses of the lines correspond to varying the absolute value of the quartic coupling $|\lambda_3|$ between 0 (darkest lines) and 2 (lightest lines) in intervals of 0.4 (note that for $|\lambda_3|=0$ the quartic and interference terms vanish).}
\label{fig:lambda3dependence}
\end{center}
\end{figure}

\begin{figure}
\includegraphics[width=0.32\textwidth]{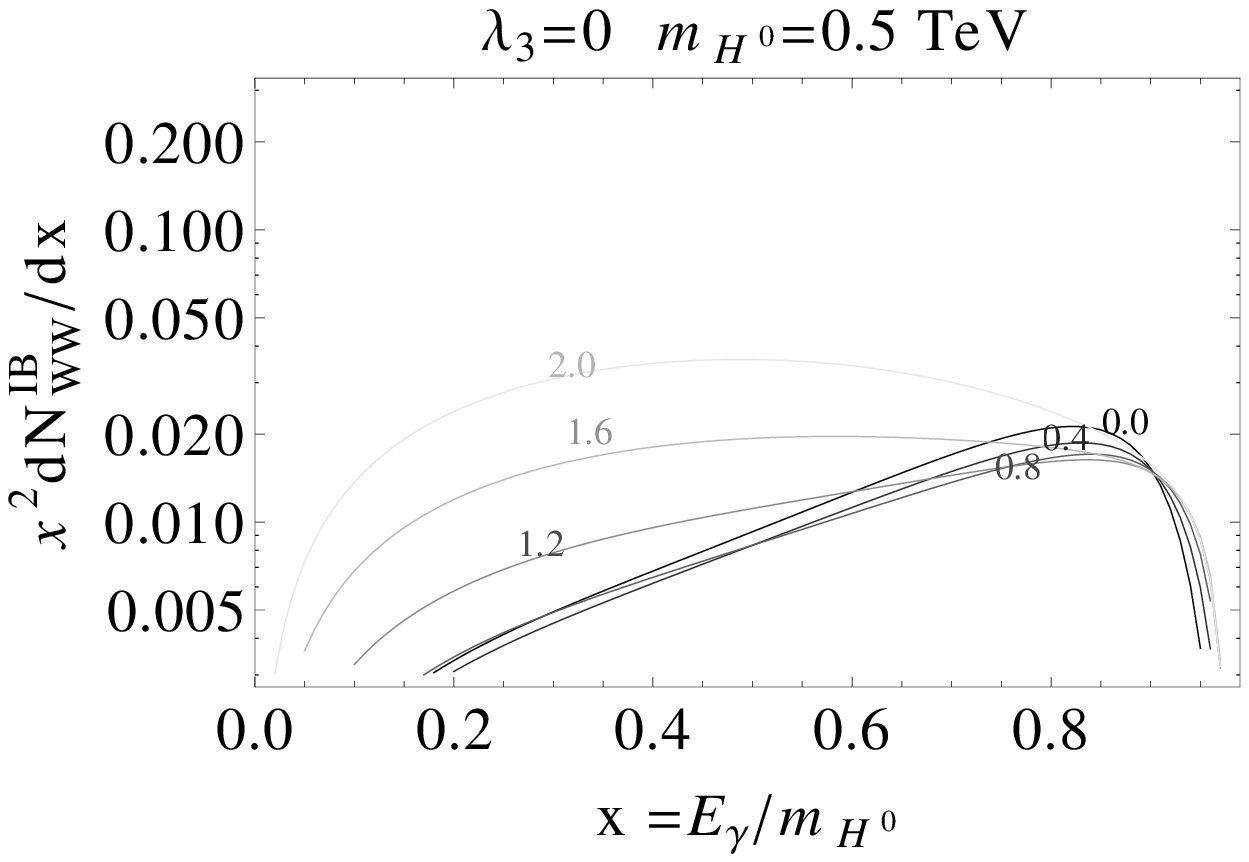}
\includegraphics[width=0.32\textwidth]{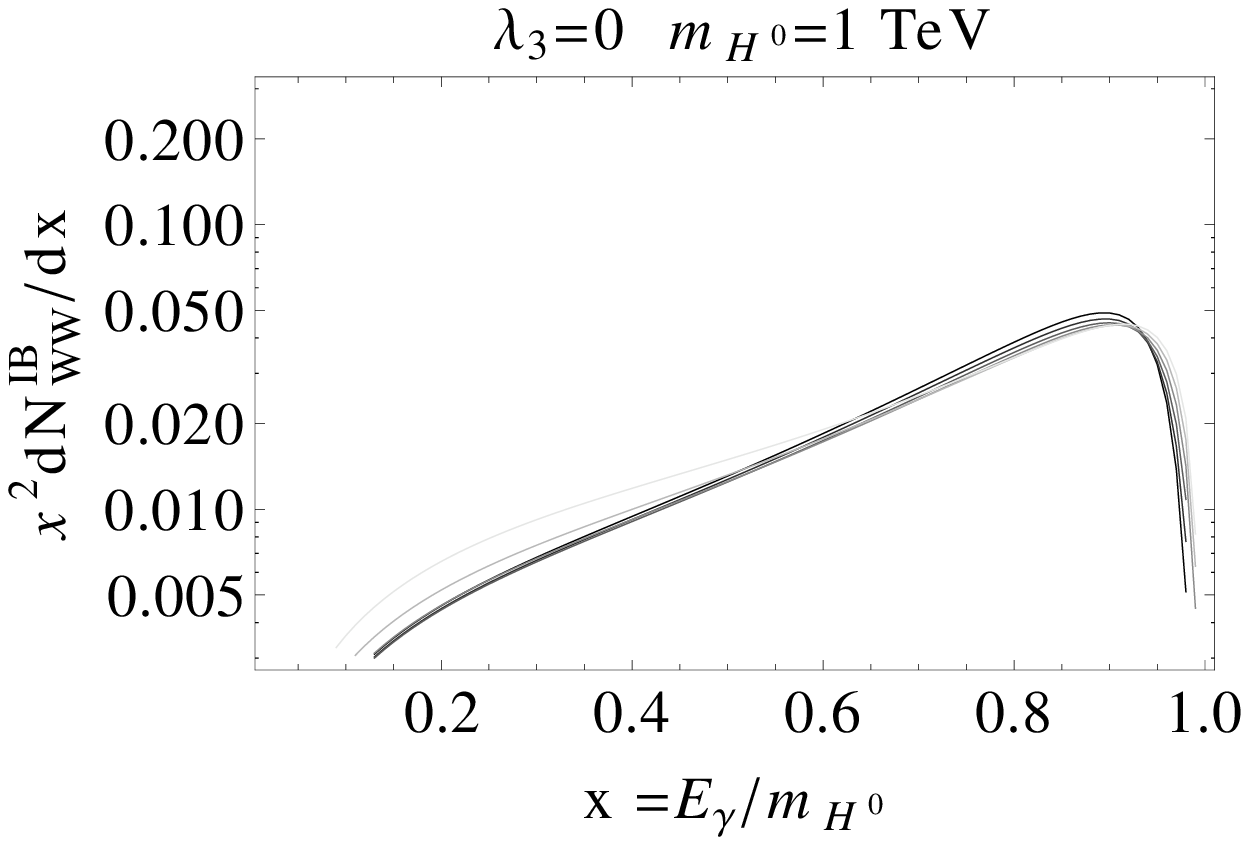}
\includegraphics[width=0.32\textwidth]{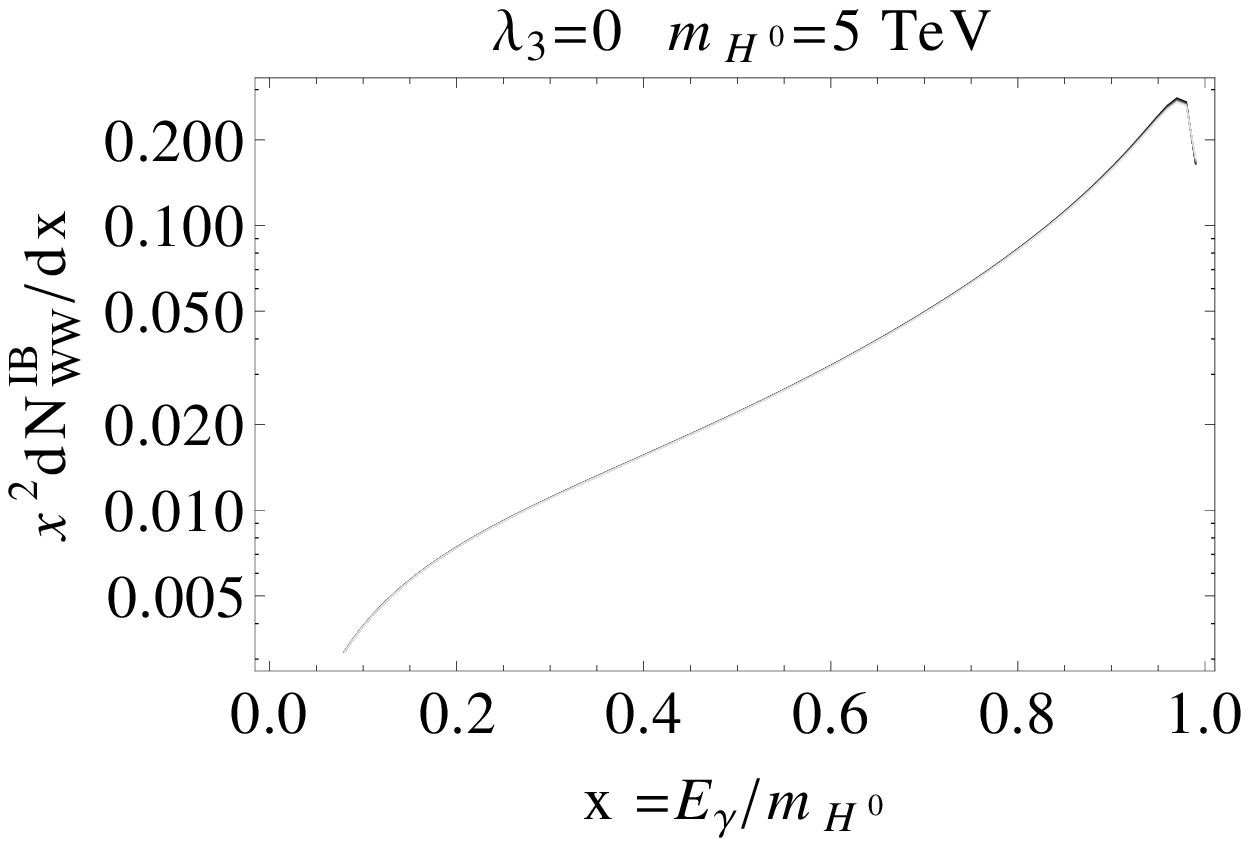}
\caption{\footnotesize Total multiplicity of the hard photon produced in the internal bremsstrahlung process $H^0 H^0\rightarrow W^+W^-\gamma$ in the presence of a mass splitting between the dark matter mass and the charged Higgs when $m_{H^0}=0.5$ TeV (left plot), 1 TeV (middle plot) and 5 TeV (right plot). In the plots $\lambda_3$ has been fixed to zero while $(\lambda_4+\lambda_5)/2$, which following Eq.~(\ref{eq:mass-splitting}) controls the mass splitting, varies between 0 (darkest lines) and $-2$ (lightest lines) in intervals of 0.4.
}
\label{fig:splittingdependence}
\end{figure}

For completeness, we also analyze the photon multiplicity from internal bremsstrahlung when the neutral and charged exotic Higgs particles are not degenerate in mass. The result is shown in Fig.~\ref{fig:splittingdependence}, where we fixed  $\lambda_3=0$ and we changed $\lambda_{H^0}=(\lambda_4+\lambda_5)/2$ from 0 to $-2$ in intervals of $-0.4$, from darkest to lightest; the mass splitting corresponding to that choice of quartic couplings can be easily derived from
\begin{equation}
M^2_{H^+}-M^2_{H^0}=-(\lambda_4+\lambda_5)\langle \Phi^0\rangle^2\;.
\label{eq:mass-splitting}
\end{equation}
Remarkably, the spectrum is quite insensitive to the mass splitting, as apparent from the plot, especially for large dark matter masses. This behaviour can be understood from expanding the total amplitude squared in the mass splitting:
\begin{equation}
|{\cal M}(\lambda_3,M_{H^0},M_{H^+})|^2 = |{\cal M}(\lambda_3,M_{H^0},M_{H^0})|^2 + {\cal O}\left[\left(\frac{M_W^2}{M_{H^0}^2}\right) \left(\frac{M^2_{H^+}-M^2_{H^0}}{\langle \Phi^0\rangle^2}\right)\right]\;.
\end{equation}
Since $(M^2_{H^+}-M^2_{H^0})/\langle \Phi^0\rangle^2$ is of order one according to eq.~(\ref{eq:mass-splitting}), we find an additional suppression in the correction term by $M_W^2/M_{H^0}^2$. Therefore, in the limit $M_{H^0}\gg M_W$ the cross section in the non-degenerate case can be safely approximated by the cross section in the degenerate case.

\section{Gamma-ray signals}
\label{sec:gamma-signals}

\begin{figure}[t]
\begin{center}
\includegraphics[width=0.49\textwidth]{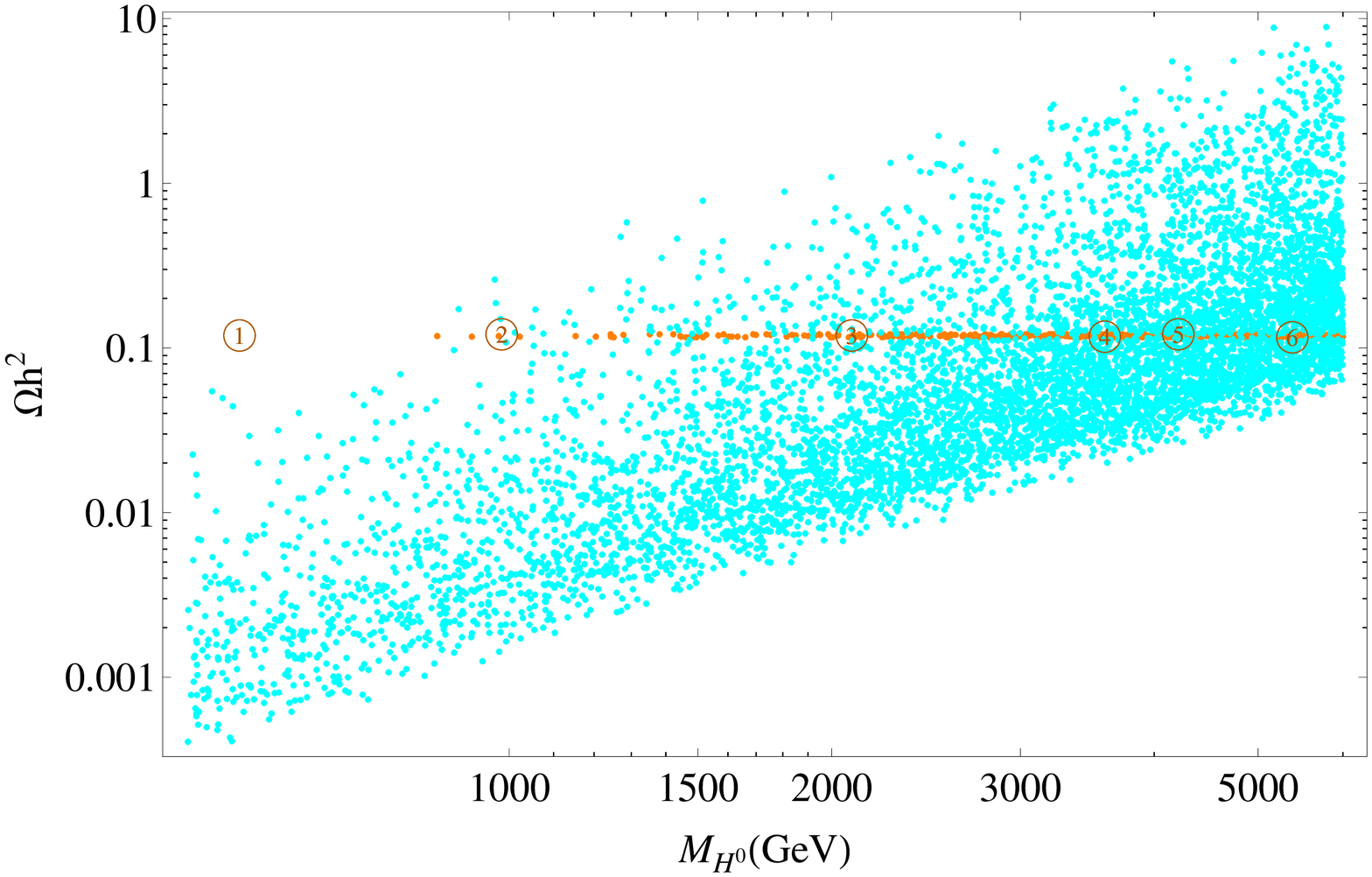} \\
\includegraphics[width=0.49\textwidth]{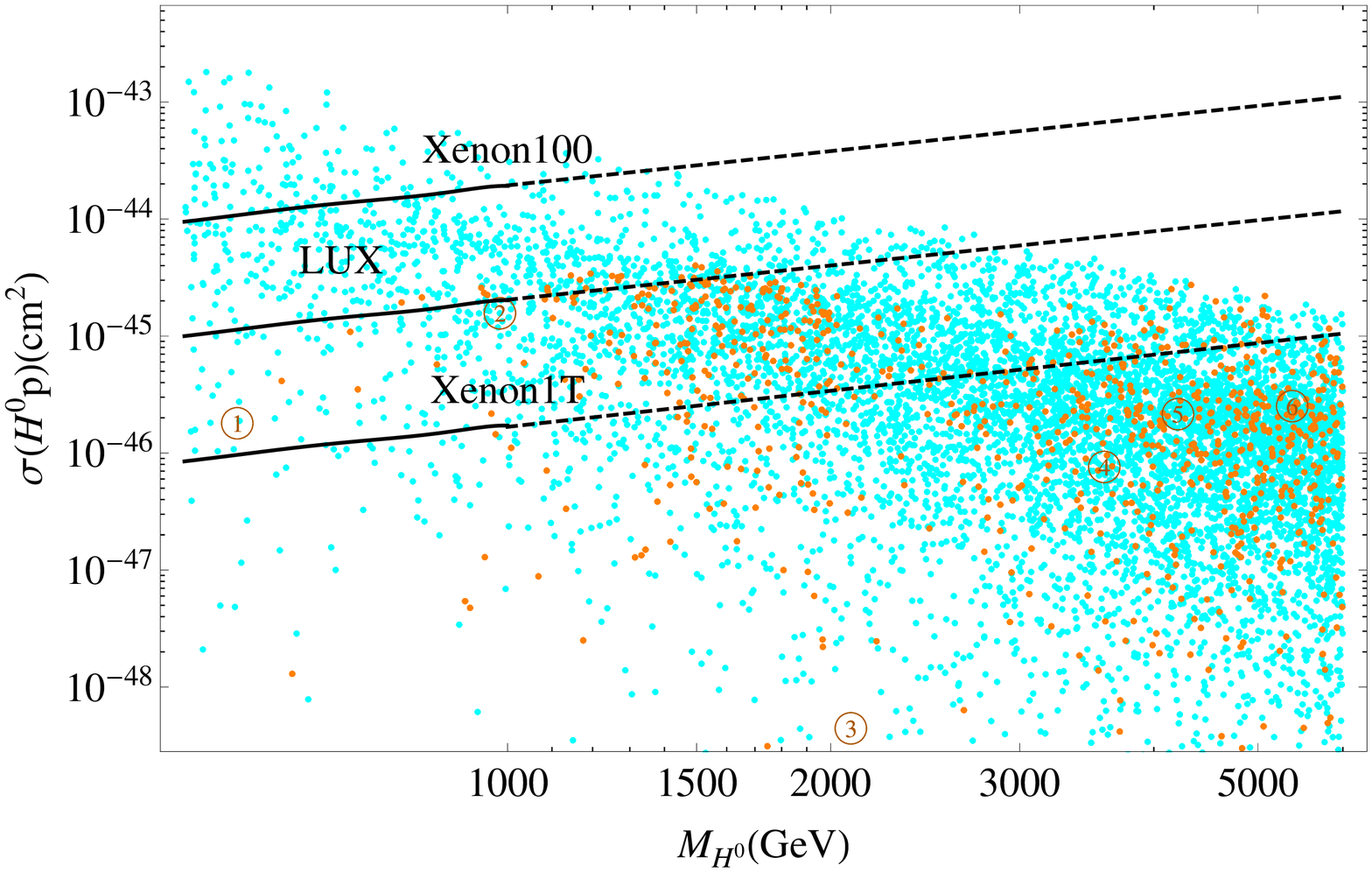}  
\includegraphics[width=0.49\textwidth]{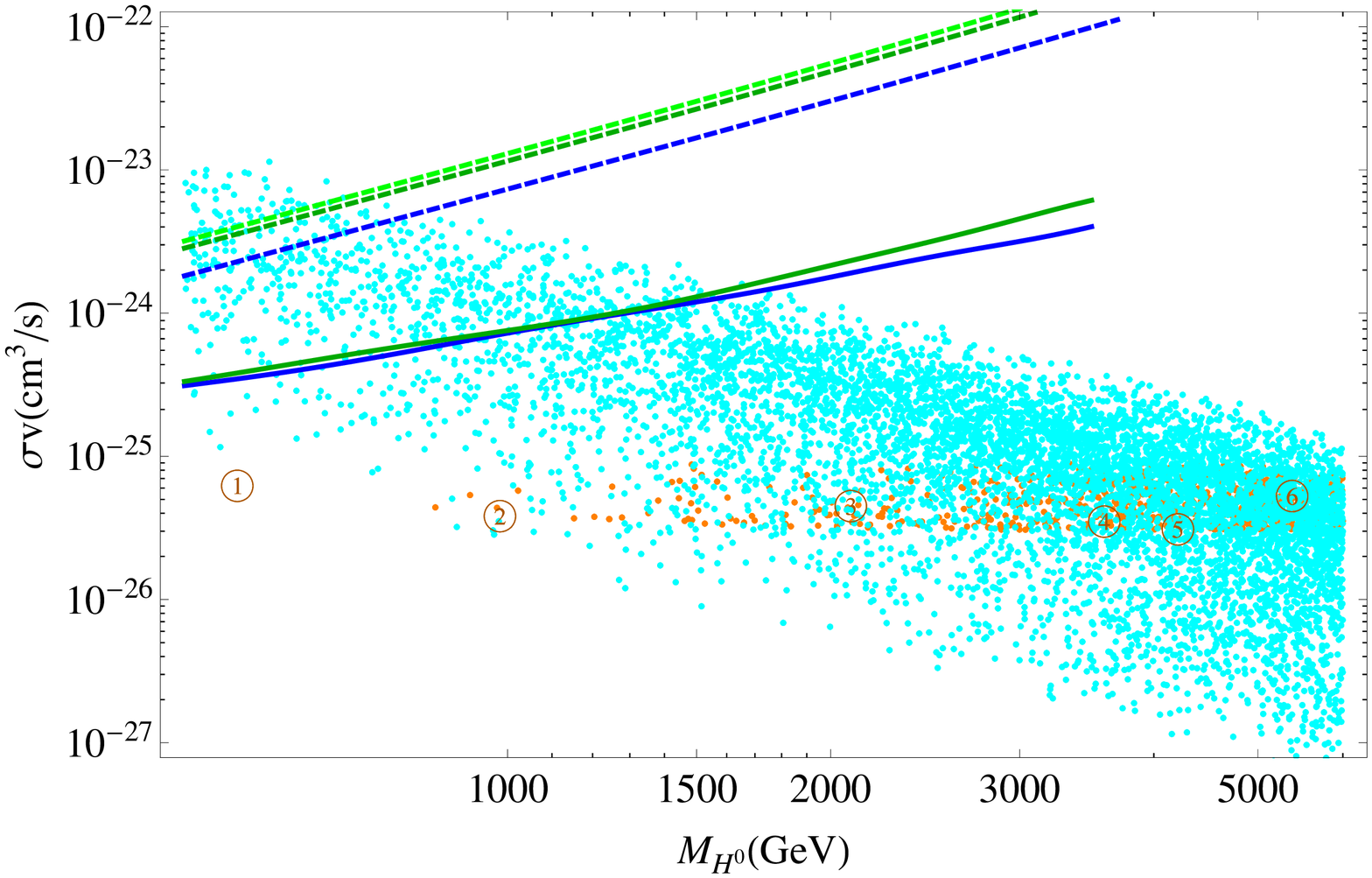} \\
\caption{\footnotesize Relic abundance (upper plot), scattering cross section with protons (bottom left plot), and total annihilation cross section (bottom right plot) obtained from scanning the parameter space of the inert doublet model. The orange points correspond to those choices that correctly reproduce the cold dark matter relic abundance and we highlight with numbers our six benchmark points in tables \ref{table:parameters}, \ref{table:xsections}. We show, together with the predicted scattering cross sections, the limits from XENON100, as well as the projected sensitivity of XENON1T and LUX. Besides, we show together with the total annihilation cross sections the limits from dwarf galaxies (dashed lines) and antiprotons (solid lines) assuming 100\% branching fraction into $WW$ (green lines) or $hh$ (blue lines). 
}
\label{fig:scatterplots}
\end{center}
\end{figure}

To study the prospects to observe the signal from internal bremsstrahlung in gamma-ray telescopes we have constructed a series of benchmark points defined by the parameters in table \ref{table:parameters}. All these points correctly reproduce the observed cold dark matter density, as illustrated in Fig.~\ref{fig:scatterplots}, upper plot, which shows the predicted dark matter relic density for various points of the parameter space of the IDM calculated using MicrOmegas3.1~\cite{Belanger:2013oya}, working under an implementation of the inert doublet model made with FeynRules~\cite{Christensen:2008py}.  The points in cyan were generated by letting the quartic couplings vary linearly in the range $|\lambda_i|\lesssim 2$; the points in orange are the subset of those points which reproduce the correct relic density, and among them we highlight our benchmark scenarios of the table.\footnote{Dark matter masses larger than 6 TeV are possible in the inert doublet model, provided $|\lambda_i|\gtrsim 2$. As discussed in Section \ref{sec:IB}, larger quartic couplings imply a larger contribution from the final state radiation to the process $H^0 H^0\rightarrow W^+ W^- \gamma$, which erases the sharp feature. Therefore, we will restrict in this paper to $|\lambda_i|\lesssim 2$ which implies $M_{H^0}\lesssim 6$ TeV.}

\begin{table}
\centering
\begin{tabular}{|c|c|c|c|c|c|c|c|c|}\hline
 BMP & $M_{H^0}\text{(GeV)}$ & $M_{H^+}\text{(GeV)}$ & $M_{A^0}\text{(GeV)}$ & $\lambda _3$ & $\lambda _4$ & $\lambda _5$ & $\lambda _{H^0}$ & $\lambda _{A^0}$ \\\hline
 1 & 559.99 & 561.85 & 560.67 & $-$0.02 & $-$0.06 & $-$0.01 & $-$0.05 & $-$0.03 \\\hline
 2 & 983.75 & 993.60 & 991.79 & 0.17 & $-$0.38 & $-$0.26 & $-$0.23 & 0.03 \\\hline
 3 & 2088.3 & 2090.99 & 2100.26 & 0.39 & 0.46 & $-$0.83 & 0.01 & 0.83 \\\hline
 4 & 3596.67 & 3597.99 & 3609.7 & $-$0.07 & 1.24 & $-$1.55 & $-$0.19 & 1.36 \\\hline
 5 & 4212.49 & 4213.09 & 4225.29 & $-$0.58 & 1.61 & $-$1.78 & $-$0.37 & 1.41 \\\hline
 6 & 5382.08 & 5382.21 & 5392.59 & 1.08 & 1.82 & $-$1.87 & 0.51 & 2.38 \\\hline
\end{tabular}
\caption{\footnotesize Dark matter ($H^0$), charged Higgs ($H^+$) and pseudoscalar ($A^0$) masses, as well as the Lagrangian parameters $\lambda_{3,4,5}$ and the quartic couplings $\lambda_{H^0}$ and $\lambda_{A^0}$ defined in Eq.~(\ref{masses}) for our six benchmark points.}
\label{table:parameters}
\end{table}

The inert Higgs doublet model is subject to constraints from direct and indirect dark matter searches. Assuming that the WIMP-nucleon scattering is dominated by Higgs exchange, the spin-independent cross-section can be approximated by ~\cite{Barbieri:2006dq}:
\begin{equation}
\sigma_{H^0-p}\simeq \frac{f^2 \lambda_{H^0}^2}{\pi}\left(\frac{m_p^2}{M_{H^0} m_h^2}\right)^2\simeq 5\times 10^{-44}\lambda_{H^0}^2 \left(\frac{1\,\TeV}{M_{H^0}}\right)^2 \,\cm^2
\end{equation}
where $m_p$ is the nucleon mass, $f\simeq 0.3$ is the nucleonic matrix element and $\lambda_{H^0}$ was defined in Eq.(\ref{masses}). We quote in table \ref{table:xsections} the spin-independent WIMP-nucleon cross-sections for our benchmark points calculated with MicrOmegas3.1, which are, as shown in Fig.~\ref{fig:scatterplots}, bottom left plot, below the present sensitivity of the XENON100 experiment \cite{Aprile:2012nq}; some of the points are, interestingly, within the sensitivity of the LUX~\cite{Akerib:2012ys} and XENON1T~\cite{Aprile:2012zx} experiments. We also quote in the table the annihilation cross section in the dominant $2\rightarrow 2$ annihilation channels, $H^0 H^0\rightarrow W^+W^-, Z^0 Z^0, hh , t\bar t$, and which are safely below the limits on the cross section derived in \cite{Cirelli:2013hv} in each of these channels from the PAMELA data on the cosmic antiproton-to-proton fraction \cite{Adriani:2010rc}. These annihilation channels are further constrained by gamma-ray observations of dwarf galaxies. The flux upper limit derived in \cite{GeringerSameth:2011iw} based on the joint analysis of seven Milky Way dwarfs can be translated into an approximate limit of the total annihilation cross-section $(\sigma v)\lesssim  5.0 \times 10^{-30} \text{cm}^3 ~ \text{s}^{-1} ~\text{GeV}^{-2} (8 \pi M_{\rm DM}^2 /N^{\text{tot}}_\gamma)$~\cite{Bringmann:2012vr}, where $N^{\text{tot}}_\gamma$ is the number of photons that are produced per annihilation with energies between $1$ and $100~\GeV$ and which we calculate for a given dark matter mass and annihilation channel using PYTHIA 6.4~\cite{Sjostrand:2006za}. Again, we find that all our benchmark points are below the upper limits of all the annihilation channels, as shown in Fig.~\ref{fig:scatterplots}, bottom right plot.

\begin{table}
\centering
\begin{tabular}{|c|c|c|c|c|c|c|c|c|}\hline
\multirow{2}{*}{BMP} 
& $\sigma(H^0p)$
&  $\sigma v$
& \multicolumn{5}{c|}{BR(\%)} 
& \multirow{2}{*}{$\Omega h^2$}
\\\cline{4-8}
&$(10^{-48}\text{cm}^2$) & $ (10^{-26}\text{cm}^3/\text{s})$ &  $W^+W^-$ & $ZZ$ & $hh$ & $t\bar{t}$ & $W^+W^-\gamma$ & 
\\\hline
 1 &  179.70 &6.41 & 50.21 & 43.50 &  2.57 & 0.60 & 3.03 & 0.120 \\\hline
 2 & 1565.57 &3.95 & 34.87 & 22.84 & 35.97 & 3.03 & 3.17 & 0.122 \\\hline
 3 &    0.44 &4.57 & 13.78 & 84.24 &  0.01 & 0.00 & 1.84 & 0.120 \\\hline
 4 &   76.09 &3.52 &  2.36 & 95.23 &  1.83 & 0.01 & 0.51 & 0.119 \\\hline
 5 &  215.06 &3.14 &  8.75 & 83.83 &  5.80 & 0.03 & 1.60 & 0.117 \\\hline
 6 &  251.71 &5.38 &  9.29 & 84.92 &  3.92 & 0.01 & 1.86 & 0.111 \\\hline
\end{tabular}
\caption{\footnotesize Scattering cross section with protons, total annihilation cross section, annihilation branching fractions and relic density in our six benchmark points.}
\label{table:xsections}
\end{table}

\begin{figure}[t]
\begin{center}
\includegraphics[width=0.48\textwidth]{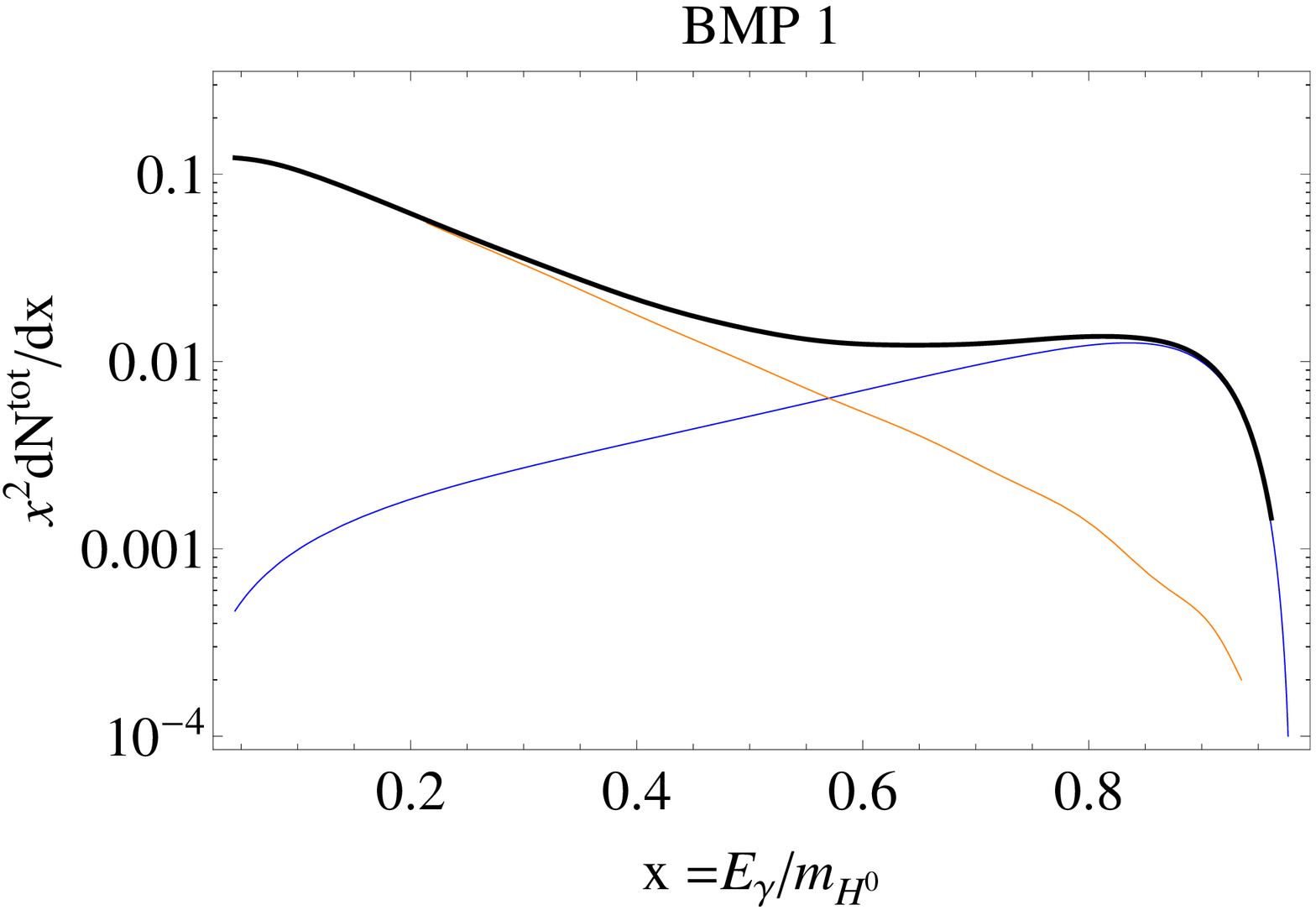}
\includegraphics[width=0.48\textwidth]{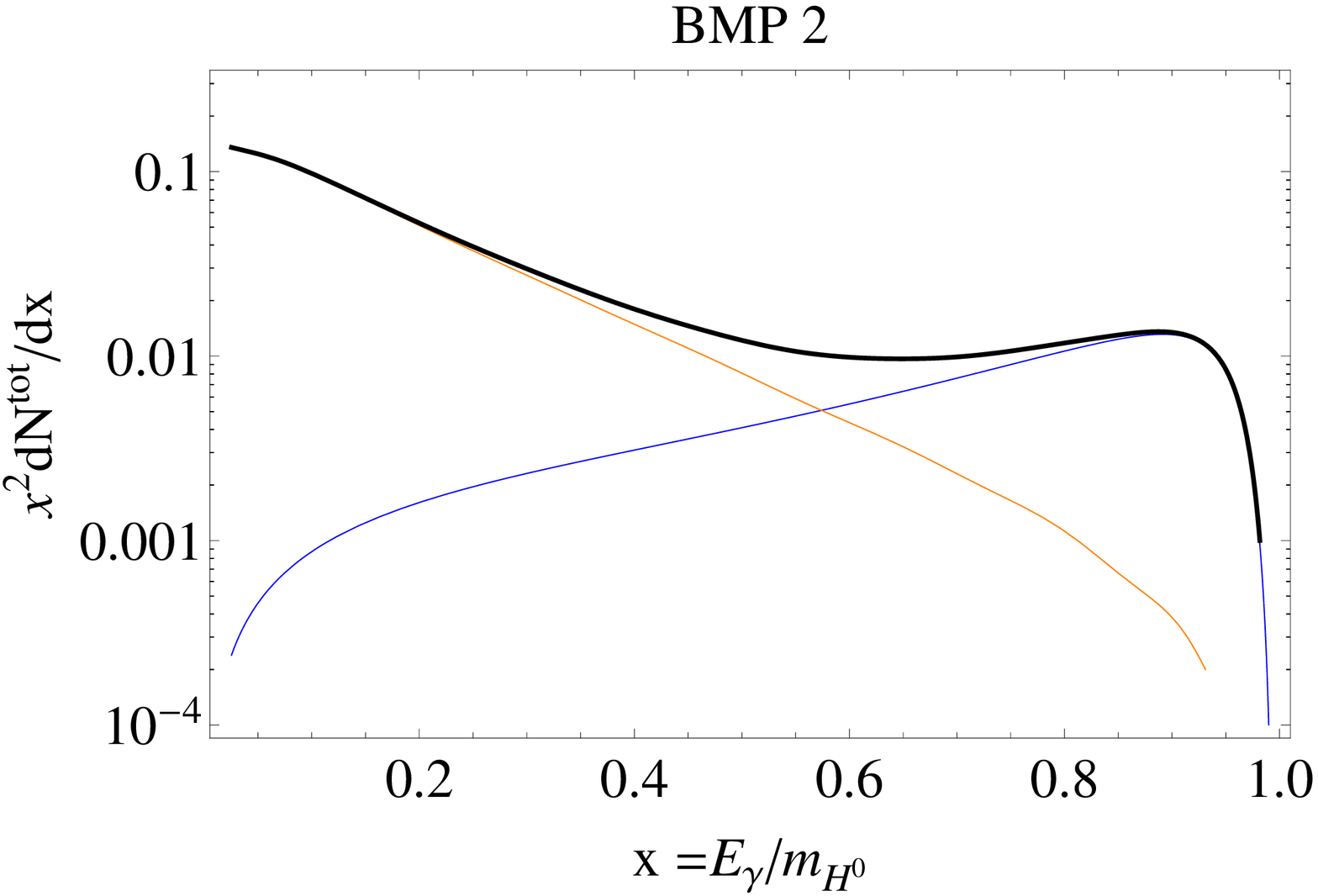}\\
\includegraphics[width=0.48\textwidth]{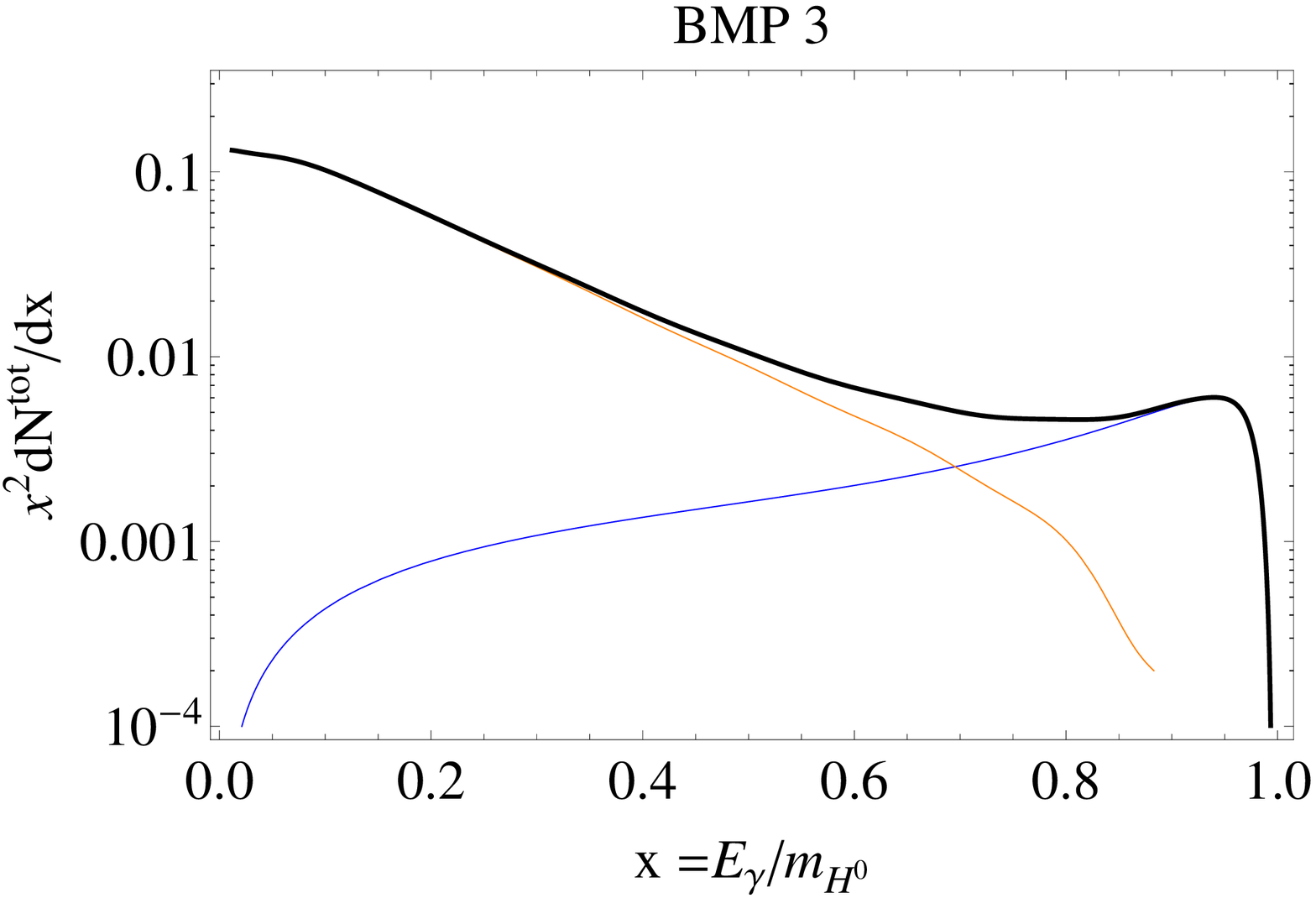}
\includegraphics[width=0.48\textwidth]{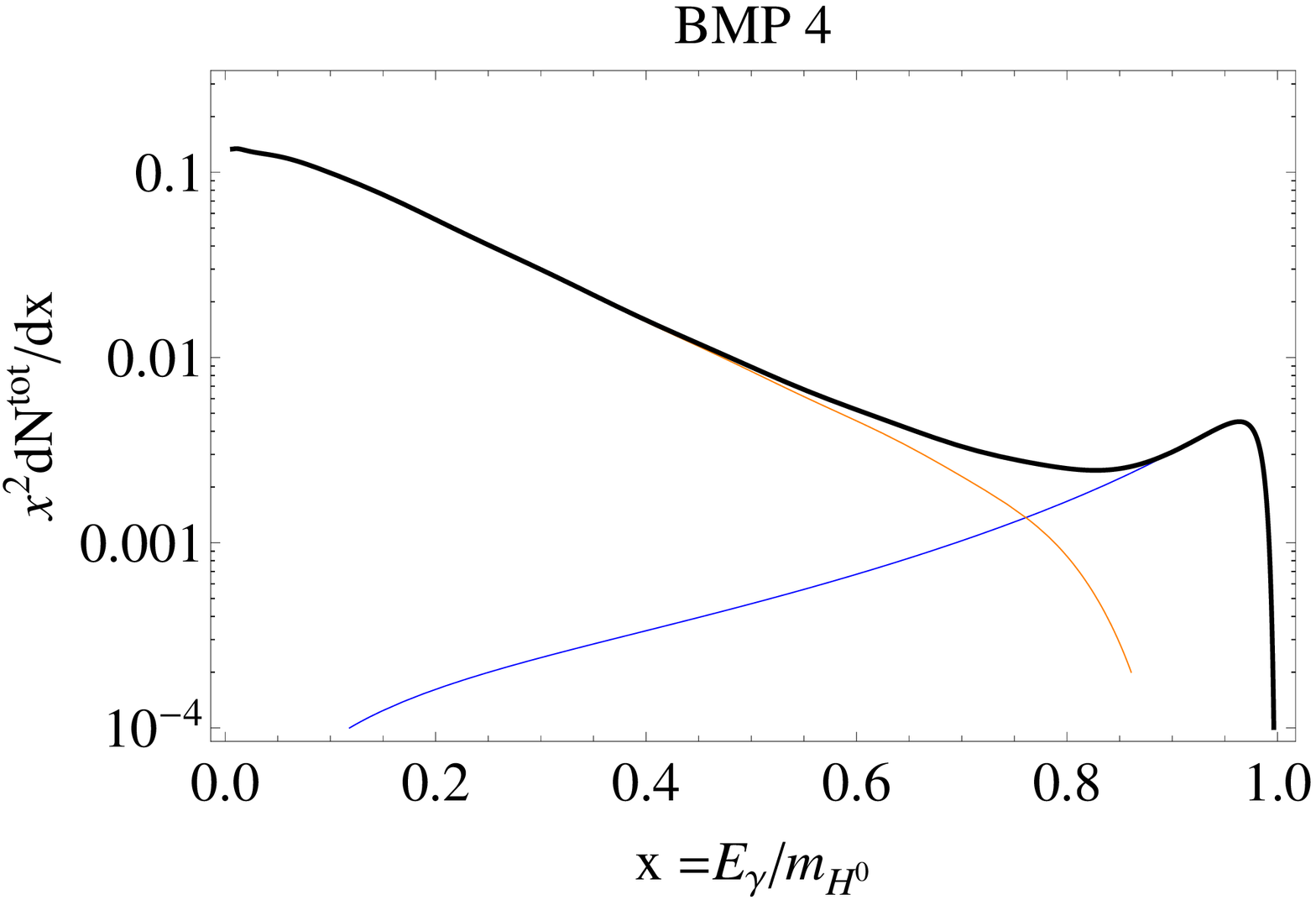}\\
\includegraphics[width=0.48\textwidth]{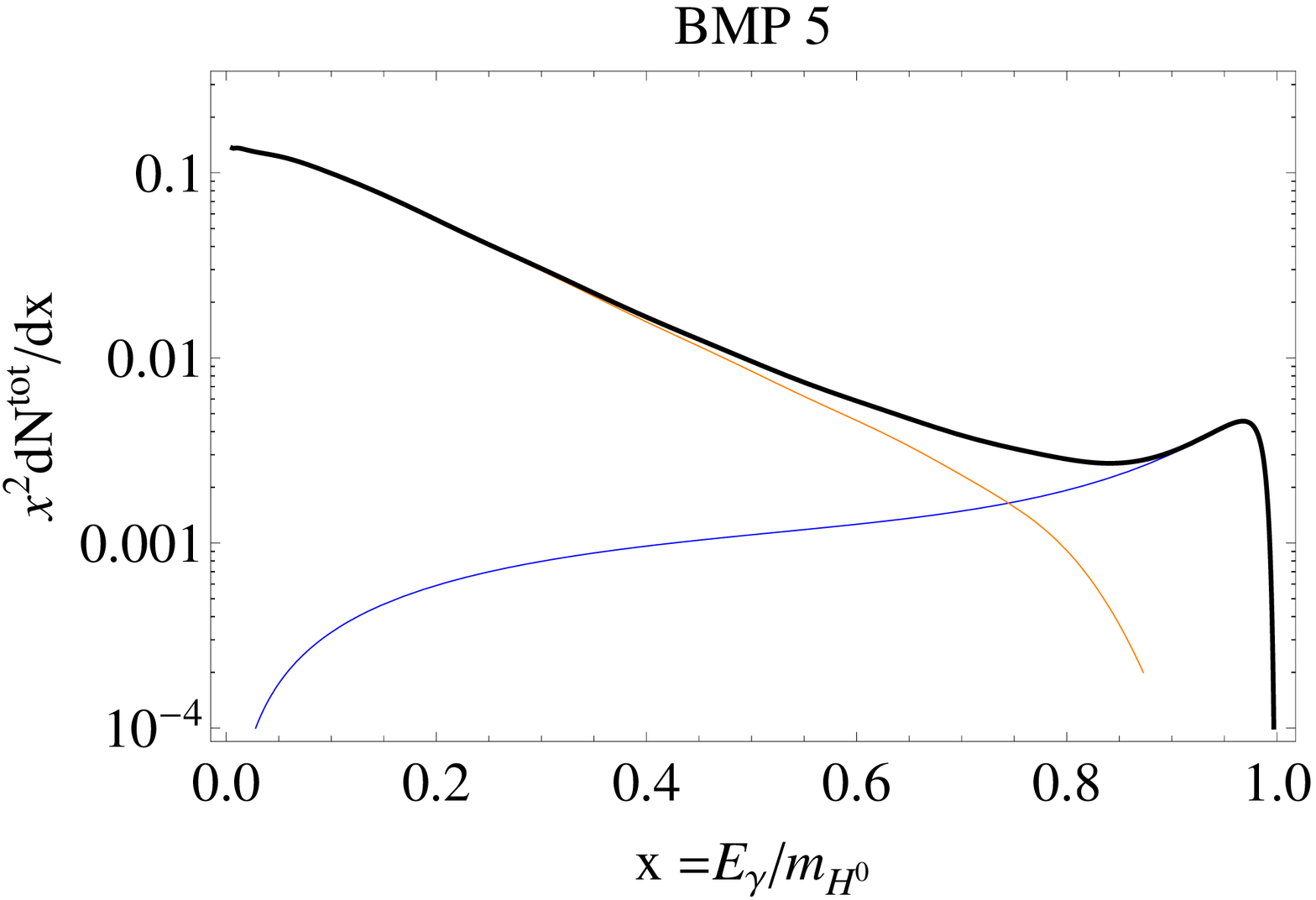}
\includegraphics[width=0.48\textwidth]{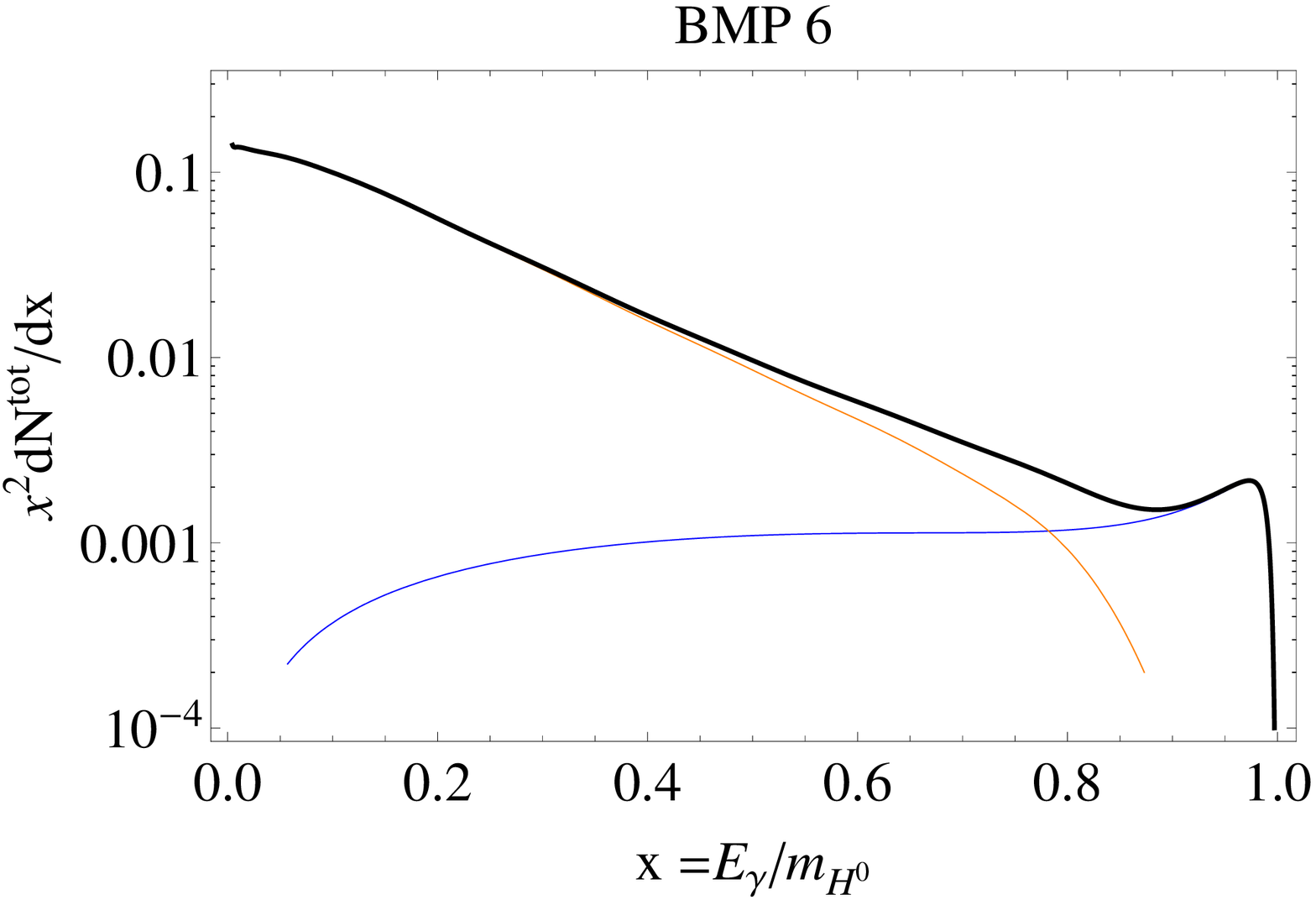}\\
\caption{\footnotesize  Predicted photon spectrum from dark matter annihilations in the inert doublet model for the six benchmark points in tables \ref{table:parameters}, \ref{table:xsections}. The red line corresponds to the photon spectrum from the $2\rightarrow 2$ processes $H^0 H^0\rightarrow W^+ W^-, Z^0 Z^0 , h h, t \bar{t} $, following the branching fractions in  tables \ref{table:xsections}, while the blue line to the $2\rightarrow 3$ process $H^0 H^0\rightarrow W^+ W^- \gamma$; the red line shows the total multiplicity resulting from summing up both contributions.}
\label{fig:BMPs}
\end{center}
\end{figure}

All these benchmark scenarios present a prominent gamma-ray feature close to the kinematic endpoint of the spectrum, as shown in Fig.~\ref{fig:BMPs}. In fact, the appearance of such a feature is generic in the inert doublet model, as long as the relevant quartic couplings $\lambda_3$, $\lambda_4$ and $\lambda_5$ are not too large, namely not larger than 2. Under these reasonable assumptions, the dark matter particle and the charged scalar are fairly degenerate in mass. Furthermore, as argued in Section~\ref{sec:IB} it is precisely in the scenario with degenerate masses and with quartic couplings $|\lambda_i|\lesssim 2$ when a sharp spectral feature from internal bremsstrahlung is expected to be generated.

\begin{figure}[t]
\centering
\includegraphics[width=0.49\textwidth]{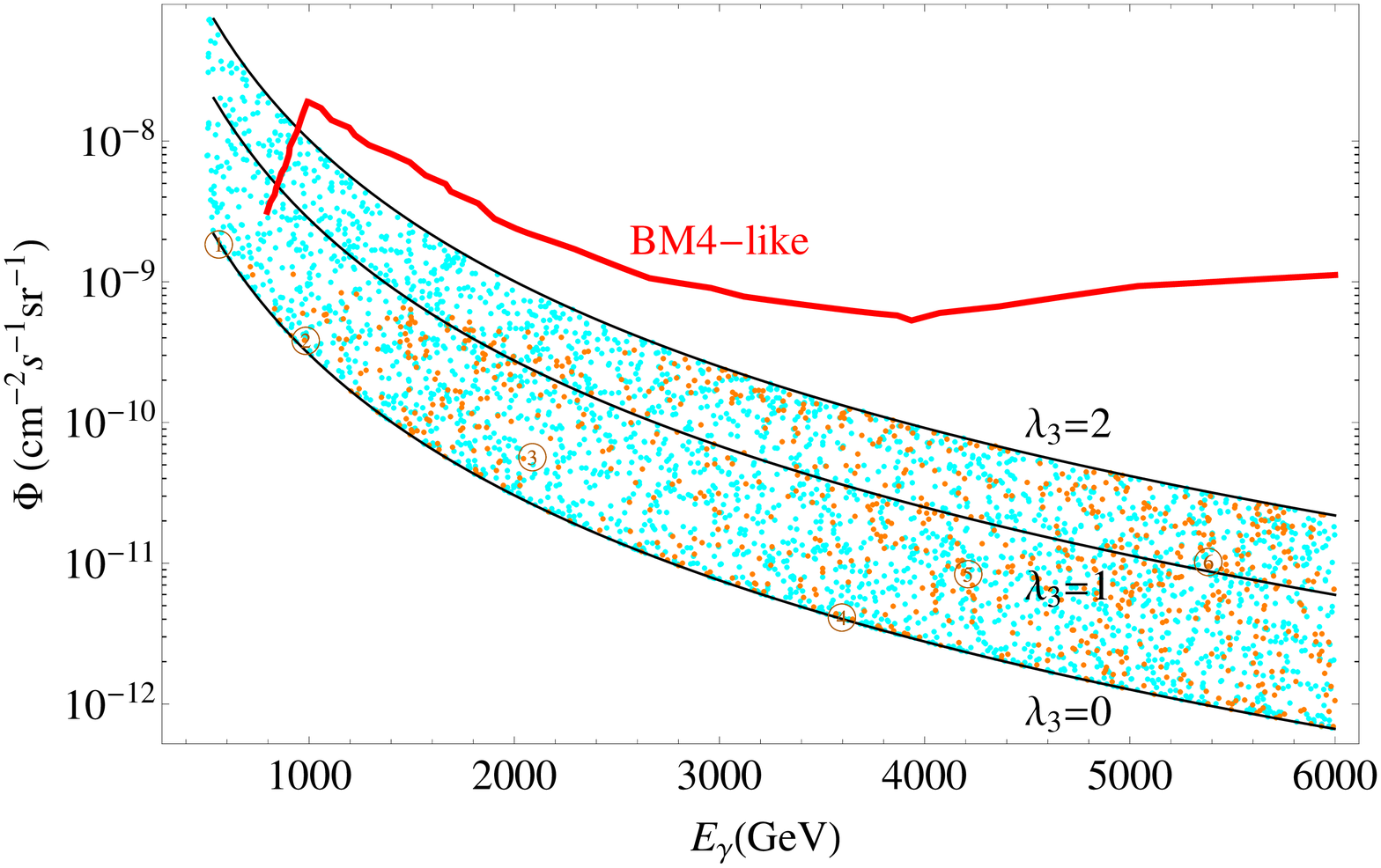}
\caption{\footnotesize Same as Fig. \ref{fig:scatterplots}, but for the integrated flux of the internal bremsstrahlung feature. We also show for reference the flux upper limit derived  in \cite{Abramowski:2013ax} by the H.E.S.S. collaboration for a ``BM4-like''spectral feature (red line) and the predicted flux calculated from the formulas in the Appendix for the case $M_{H^0}=M_{H^+}$ and $|\lambda_3|=0,~1,~2$ (black lines).}
\label{fig:HESS}
\end{figure}

The photon flux produced in dark matter annihilations and received at the Earth from a given solid angle in the sky, $\Delta \Omega$, is given by
\begin{equation}
\frac{d\phi_\gamma}{dE_\gamma } = \frac{1} {8 \pi M^2_{H^0}} 
\sum_f  (\sigma v)_f \frac{dN^{\text{tot}}}{dE_\gamma} \frac{1}{\Delta \Omega} \int_{\Delta \Omega}  J  d\Omega  \;,
\label{fluxeq}
\end{equation}
where $(\sigma v)_f$ and $\frac{dN^{\text{tot}}}{dE_\gamma}$ are, respectively, the velocity weighted annihilation cross-section and the differential energy spectrum corresponding to the annihilation channel ``$f$''. Besides, the ``$J$-factor'' is the integral of the squared dark matter density $\rho_{H^0}$ along the line of sight, $J = \int_{\text{l.o.s.}} ds \rho^2_{H^0}$. We choose for our study the Einasto profile, favoured by recent $N$-body simulations \cite{Navarro:2008kc,Hayashi:2007uk,Gao:2007gh},
\begin{equation}
\rho_{\rm DM}(r)\propto \exp\left[-\frac{2}{\alpha}\left(\frac{r}{r_s}\right)^\alpha\right]
\end{equation}
with  $r_s=20$ kpc and $\alpha=0.17$ \cite{Navarro:2003ew,Springel:2008cc}, as well as the Navarro-Frenk-White (NFW) profile~\cite{Navarro:1995iw,Navarro:1996gj}:
\begin{equation}
  \rho_\text{DM}(r)\propto\frac{1}{(r/r_\text{s})
  [1+(r/r_\text{s})]^2}\;,
\end{equation}
with scale radius $r_s = 21 ~\rm{kpc}$~\cite{Pieri:2009je}, both profiles normalized to a local dark matter density $\rho_{\rm DM}(r=8.5\textrm{ kpc})=0.39\,{\rm GeV}\,{\rm cm}^{-3}$ \cite{Catena:2009mf,Weber:2009pt,Salucci:2010qr,Pato:2010yq}. We will compare our predicted flux to the limits recently derived by the  H.E.S.S. collaboration from a search for line-like gamma-ray features in the central part of the Milky Way halo with energies between $\sim 500$ GeV and $\sim 25$ TeV~\cite{Abramowski:2013ax}, which adopts a complicated search region with a  $J$-factor given by $J=1.2\times 10^{25} \GeV^2\,\cm^{-5}$ for the Einasto profile and $J= 6.4\times 10^{24}  \GeV^2\,\cm^{-5}$ for the NFW profile ~\cite{Abramowski:2011hc}. We show in Fig.~\ref{fig:HESS} the integrated flux of the hard photon emitted in the internal bremsstrahlung process  $H^0 H^0\rightarrow W^+ W^- \gamma$ compared to the limits derived by the H.E.S.S. collaboration for the BM4 benchmark point of \cite{Bringmann:2007nk}, corresponding to the neutralino annihilation $\chi^0\chi^0\rightarrow W^+ W^- \gamma$ ~\cite{Bergstrom:2005ss} and which produces a similar spectrum as  $H^0 H^0\rightarrow W^+ W^- \gamma$. In the plot it was assumed the Einasto profile; the fluxes for the NFW profile are a factor of two smaller. As apparent from the plot, present instruments are not sensitive enough to observe the signal from internal bremsstrahlung of the inert doublet model, unless the annihilation signal is boosted by astrophysical or particle physics effects by a factor ${\cal O}(10-100)$. Future instruments, such as DAMPE~\cite{dampe}, GAMMA-400~\cite{Galper:2012fp} or CTA~\cite{Consortium:2010bc} will, however, close in on the signal of internal bremsstrahlung from the inert doublet model.

\section{Conclusions}
\label{sec:conclusions}
We have shown that the spectral feature from internal bremsstrahlung  naturally  emerges in the high mass regime of the inert doublet model, namely when $M_{H^0}\gtrsim 500\GeV$. In this regime, the higher order annihilation $H^0 H^0\rightarrow W^+ W^- \gamma$ generates, due to the mass degeneracy between the charged Higgs and the dark matter particle, a distinctive hardening in the gamma-ray energy spectrum which could be observed over the featureless gamma-ray background. We have also constructed a series of benchmark points, compatible with all theoretical and experimental constraints on the inert doublet model, which present an internal bremsstrahlung feature with an intensity which is one or two order of magnitude below the H.E.S.S. limits on this class of photon line-like signatures. Future searches for gamma-ray spectral features by DAMPE, GAMMA-400 or CTA will continue closing in on the parameter space of the inert doublet dark matter model.

\FloatBarrier

\section*{Acknowledgements}
We are grateful to Michael Gustafsson, Laura Lopez Honorez, Michel Tytgat and Stefan Vogl for useful discussions.  This work was partially supported by the DFG cluster of excellence ``Origin and Structure of the Universe'' and by the Graduiertenkolleg ``Particle Physics at the Energy Frontier of New Phenomena''. AI thanks the Kavli Institute for Theoretical Physics in Santa Barbara, California, for hospitality during the final stages of this work.

\section*{Appendix}

The differential cross-section for the process $H^0 H^0 \to W^+W^-\gamma$ can be written as
\begin{eqnarray}
d\sigma &=& \frac{1}{(2\pi)^5 J} |{\cal M}|^2 d_3(PS),
\end{eqnarray}
where ${\cal M}$ is the scattering amplitude, $J$ the initial flux and $d_3(PS)$ the three-body phase-space factor. These are defined by
\begin{eqnarray}
J&=& 4E_1 E_2 v, \hspace{30pt} \text{where  } v=\left|\frac{\vec{p_1}}{E_1}-\frac{\vec{p_2}}{E_2}\right|\;, \\ 
d_3(PS) &=&\delta^4 (p_1+p_2- p_{W^+}-p_{W^-}-p_{\gamma})   \frac{d^3p_{W^+}}{2E_{W^+}} \frac{d^3p_{W^-}}{2E_{W^-}}\frac{d^3p_{\gamma}}{2E_{\gamma}}\;.
\end{eqnarray}
Here the labelling of the momenta is self-explanatory. We are interested in the limit of zero dark matter velocity, for which $E_1,E_2 \to M_{H^0}$. Besides, the three-body phase-space factor after integration of the delta function can be written as
\begin{eqnarray}  
d_3(PS) &=& \pi^2 M^2_{H^0} dx_+ dx, \hspace{30pt} \text{where  } \hspace{30pt}  x_+ = \frac{E_{W^+}}{M_H^0}\text{     and       } x = \frac{E_\gamma}{M_H^0}\;.
\end{eqnarray}
The ranges for $x_+$ and $x$ are determined by the energy range of the photon and the $W^+$ and are given by
\begin{eqnarray}
0 < x < 1 - 4\mu, \hspace{50pt}
x_+ \lessgtr \frac{1}{2} \left(2-x \pm x \sqrt{1-\frac{4\mu}{1-x}} \right)
\end{eqnarray}
where $\mu = \frac{M_W^2}{s} = \frac{M_W^2}{4M_{H^0}^2}$. As a result we have
\begin{equation}
\frac{d(\sigma v)}{dx} \Bigg|_{v\to0}  = 
\frac{1}{128\pi^3} \int^{x_{+\text{max}}}_{x_{+\text{min}}}|{\cal M}|^2 \Bigg|_{v\to0} dx_+
\end{equation}
The diagrams that contribute to ${\cal M}$ in the unitary gauge are shown in Fig. \ref{fig:diagramswwg}. The first twelve depend on $M_{H^+}$ and $M_{H^0}$ and are proportional to $g^2 e$, where $g$ and $e$ are the weak and the electromagnetic coupling constants respectively. On the other hand, the last two diagrams depend only on $M_{H^0}$ and are proportional to $g \lambda_H e$, where $\lambda_H$ is the quartic coupling entering in the $H^0 H^0 h$ vertex. Following Eqs.(\ref{masses}) and (\ref{masssplitting}) we can write the scattering amplitude only as a function of $M_{H^0}$, $\lambda_3$, $\lambda_4$ and $\lambda_5$. Furthermore, we notice that ${\cal M}$  depends on $\lambda_4$ and $\lambda_5$ only through the combination $\lambda_4+\lambda_5$.

In order to simplify the expression for the scattering amplitude, we find useful to write the amplitude in the following way. Since gauge transformations do not depend on the quartic couplings,  ${\cal M}$ can be separated in two gauge-invariant pieces
\begin{equation}
{\cal M}(M_{H^0},\lambda_3,\lambda_4+\lambda_5)  =  {\cal M}_{\text{Gauge}} (M_{H^0}) + {\cal M}_{\text{Quartic}} (M_{H^0},\lambda_3,\lambda_4+\lambda_5)
\end{equation}
where the first term is the scattering amplitude when the quartic couplings are set to zero. Accordingly, the squared amplitude can be cast as 
\begin{equation}
|{\cal M}|^2 =|{\cal M}_{\text{Gauge}}|^2 +|{\cal M}_{\text{Quartic}} |^2+2 Re  \left( {\cal M}_{\text{Gauge}}{\cal M}_{\text{Quartic}}^\dagger \right)
\end{equation}
The expression for $|{\cal M}|^2$ is very complicated in general. However, in the limit of zero dark matter velocity, we find that
\begin{equation}
|{\cal M}(M_{H^0},\lambda_3,\lambda_4+\lambda_5)|^2 \Bigg|_{v\to0} = |{\cal M}(M_{H^0},\lambda_3,0)|^2 \Bigg|_{v\to0} + O(\mu(\lambda_4+\lambda_5)) \;,
\end{equation}
since we expect $\mu$ to be very small, the dependence on $\lambda_4+\lambda_5$ is subdominant. As a result, it is a good approximation to take $\lambda_4+\lambda_5 \approx 0$  (or $M_{H^0} \approx M_{H^+}$), especially for dark matter masses much heavier than the $W$ mass. Under that approximation and dropping the label $v \to 0$, we find
\begin{align}
%
|{\cal M}_{\text{Gauge}}|^2 =& \frac{e^2 g^4}{2M_{H^0}^2 (x_+-1)^2 (x+x_+-1)^2 (x_+-2 \mu )^2 (2\mu +x+x_+-2)^2}
\nonumber\\
&\hspace{-1cm}\Bigg[ x^6 \Big(-\mu +x_+-1\Big) 
+x^5 \Big(3 x_+^2+(2 \mu -9) x_+ -4 \mu ^2+6\Big)
+x^4 \Big(5 x_+^3-(\mu +20) x_+^2
\nonumber\\
&\hspace{-1cm}+2 (7 \mu ^2-6 \mu +15)x_+ -10 \mu ^3-6 \mu ^2+11 \mu -15\Big)  
+x^3 \Big(5 x_+^4-(6 \mu +25) x_+^3 
\nonumber\\
&\hspace{-1cm}+2 (3 \mu ^2+6 \mu +25) x_+^2+ (40 \mu ^3-66 \mu ^2+20 \mu -50) x_+-4 (6 \mu ^4+3 \mu ^3-12 \mu ^2+6 \mu -5)\Big)
\nonumber\\
&\hspace{-1cm}+x^2 \Big(3 x_+^5-(3 \mu +20) x_+^4+\left(-16 \mu ^2+28 \mu
   +50\right) x_+^3+(40 \mu ^3+2 \mu ^2-44 \mu -65) x_+^2
\nonumber\\
&\hspace{-1cm}+2 (12 \mu ^4-72 \mu ^3+49 \mu ^2-2 \mu +24)
   x_+-16 -24 \mu ^5+16 \mu ^4+78 \mu ^3-76 \mu ^2+22 \mu \Big)
\nonumber\\
&\hspace{-1cm}
+x (x_+-1) \Big(x_+^5-8 x_+^4+(-8 \mu ^2+8
   \mu +22) x_+^3+4 (10 \mu ^2-8 \mu -7) x_+^2 
\nonumber\\
&\hspace{-1cm}+(24 \mu ^4-64 \mu ^3-2 \mu ^2+20 \mu +20) x_+-4 (12 \mu ^4-26 \mu ^3+14 \mu ^2-3 \mu +2)
 \Big)
\nonumber\\
&\hspace{-1cm}-(x_+-1)^2 \Big(x_+^4-4 x_+^3+(-8 \mu ^2+4 \mu +6) x_+^2+4 (4 \mu ^2-2 \mu -1) x_+ 
\nonumber\\
&\hspace{-1cm}+ 24 \mu ^4-40 \mu ^3+18 \mu ^2-4 \mu +2 \Big) 
\Bigg],
\end{align}
%
\begin{align}
|{\cal M}_{\text{Quartic}}|^2 =& \frac{e^2\lambda_3^2}{M_{H^0}^2 (x_+-1)^2 (x+x_+-1)^2(1-\mu_h)^2}
\nonumber \\
&\hspace{-1cm}\Bigg[
2 \mu  x^4
+4 \mu  x^3 (x_+-1)+
x^2 (4 \mu  x_+^2+12 \mu ^2 x_+-12 \mu  x_++x_+-12 \mu ^3-8 \mu ^2+7 \mu -1)
\nonumber \\
&\hspace{-1cm}+x(12 \mu ^2-4 \mu +1)(x_+^2-3 x_++2)
-(12 \mu ^2-4 \mu+1) (x_+-1)^2
\Bigg],
\end{align}
%

\begin{align}
2 Re & \left( {\cal M}_{\text{Gauge}}{\cal M}_{\text{Quartic}}^\dagger \right) =
\frac{2e^2g^2 \mu  \lambda_3}{M_{H^0}^2 (x_+-1)^2 (x+x_+-1)^2 (1-\mu_h)(x_+-2 \mu ) (2 \mu +x+x_+-2)}
\nonumber \\
&\hspace{-1cm}\Bigg[x^4 (1-x_+)
+x^3 \Big(x_+^2+(3-12 \mu ) x_+ + 6 \mu ^2+9 \mu -4\Big)
+x^2 \Big(4 x_+^3-3 (4 \mu +3) x_+^2
\nonumber \\
&\hspace{-1cm}-6 \mu  (2 \mu -7) x_++12\mu ^3-27 \mu +5\Big)
+x (x_+-1) \Big(2x_+^3-10 x_+^2+2(-6 \mu ^2+9 \mu +4) x_+ 
\nonumber \\
&\hspace{-1cm}+24 \mu ^2-30 \mu +3\Big)-(x_+-1)^2 \Big(-12 \mu ^2+12 \mu +2 x_+^2-4 x_+-1\Big)
\Bigg],
\label{InterferenceAmp}
\end{align}
where $\mu_h = \frac{M_h^2}{s} = \frac{M_h^2}{4M_{H^0}^2}$. Notice that the interference term is proportional to $\mu$ and it is therefore subdominant. Under the same approximation, the corresponding parts of the velocity weighted cross-section are given by
\begin{align}
\frac{d(\sigma v)}{dx}\Bigg|_{\text{Gauge}}&=
-\frac{e^2 g^4}{128\pi^3 M_{H^0}^2  (1-2 \mu)^3} 
\\
&\hspace{-2cm}\times\left[\left(
\frac{
2-4 x+6 x^2-4 x^3+x^4-10 \mu+16 x \mu-20 x^2 \mu+8 x^3 \mu+16 \mu^2-20 x \mu^2+18 x^2 \mu^2-8 \mu^3+8 x \mu^3
}{(2-x-4 \mu)^2 \left(1-2 x+x^2-4 \mu+6 x \mu-x^2 \mu+4 \mu^2-4 x \mu^2\right)}
\right. \right. \nonumber \\ 
&\hspace{-2cm} \left.\left.
+\frac{1+x^2-4 \mu+6 \mu^2}{x^2  (1-\mu)} \right) x(1-x) (1-\mu)(1-2 \mu)A(x) 
\right. \nonumber \\ 
& \hspace{-2cm} \left.
+\frac{1}{x (1-x-2\mu)}
\Bigg(
x^4-2 x^3 (1-2 \mu) (1-\mu)+x^2 (3-4 \mu) \left(1-4 \mu+6 \mu^2\right)
\right. \nonumber \\ 
& \hspace{-2cm} \left. 
-2 x (1-2 \mu)^2 \left(1-4 \mu+6 \mu^2\right)+(1-2 \mu)^3 \left(1-4 \mu+6 \mu^2\right) \Bigg) B(x)
\right. \nonumber \\ 
& \hspace{-2cm} \left.
+\frac{1}{(2-x-4 \mu)^3 (1-x-2 \mu)}
\Bigg( 
x^6-2 x^5 (1-2\mu) (4-\mu)-2 x^4 (-12+53 \mu-71 \mu^2+24 \mu^3 )
\right. \nonumber \\ 
& \hspace{-2cm} \left.
-4 x^3 (9-57 \mu+136 \mu^2-142 \mu^3+52 \mu^4) -2 x^2 (1-2 \mu)^2 (-14+54 \mu-87 \mu^2+46 \mu^3)
\right. \nonumber \\ 
& \hspace{-2cm} \left. 
-4 x (-1+2 \mu)^3 (-3+8 \mu-13 \mu^2+6 \mu^3)+4 (1-2 \mu)^4 (1-2 \mu+3 \mu^2)
\Bigg) C(x)
\right]\,,
\label{Gauge}
\end{align}
\begin{align}
\frac{d(\sigma v)}{dx}\Bigg|_{\text{Quartic}}&= 
\frac{32 e^2 M_{H^0}^2 \lambda_3^2} {128\pi^3 \left(4 M_{H^0}^2-M_h^2\right)^2}
\nonumber \\
&\hspace{-1cm}\times \left[-\left( \frac{1-x}{x} \right) \left(1-2 x^2-4 \mu+12 \mu^2 \right)A(x) 
+\left(\frac{1-x-2 \mu}{x}\right)  \left(1-4 \mu+12 \mu^2\right) B(x) \right]
\;,
\label{Quartic}
\end{align}
\begin{align}
\frac{d(\sigma v)}{dx}\Bigg|_{\text{Interference}}&=
\frac{16 e^2 g^2 \lambda_3 \mu}{128\pi^3\left(4M_{H^0}^2-M_h^2\right)} \left[-3\left(\frac{1-x}{x}\right) A(x) 
\right. \nonumber \\ 
&\hspace{-2cm}\left.+\left(\frac{x^4-x^3 (7-13 \mu)-15 x (1-2 \mu)^3+6 (1-2 \mu)^4+2 x^2 \left(8-31 \mu+30 \mu^2\right)}{x (2-x-4 \mu) (1-2 \mu) (1-x-2 \mu)}\right) B(x)
\right. \nonumber \\ 
&\hspace{-2cm}\left.+\left( \frac{ x^3-(1-2 \mu)^2+x^2 (-3+5 \mu)+x \left(2-6 \mu+4 \mu^2\right)}{(2-x-4 \mu) (1-2 \mu) (1-x-2\mu)}\right) C(x) \right]\;,
\label{Interference}
\end{align}
where
\begin{eqnarray}
A(x) = \sqrt{1-\frac{4\mu}{1-x}}, \hspace{40pt}
B(x) &=& \log\left[\dfrac{1+A(x)}{1-A(x)}\right], \hspace{40pt}  
C(x) =\log\left[\dfrac{2-x+A(x) x-4 \mu}{2-x-A(x) x-4 \mu}\right] \;. \nonumber 
\end{eqnarray}
%


\bibliographystyle{JHEP}
\bibliography{text}

\end{document}

%% file: figure1.tex
{
\unitlength=1.0 pt
\SetScale{1.0}
\SetWidth{0.7}      
\scriptsize    
{} \qquad\allowbreak
\begin{picture}(120,62)(0,0)
\DashLine(60.0,59.0)(36.0,59.0){1.0}
\Text(36.0,59.0)[r]{$H^0$}
\DashArrowLine(60.0,59.0)(60.0,35.0){1.0} 
\Text(63.0,48.0)[l]{$H^+$}
\Photon(84.0,59.0)(60.0,59.0){2.0}{4} 
\Text(84.0,59.0)[l]{ $W^-$}
\DashArrowLine(60.0,35.0)(60.0,11.0){1.0} 
\Text(63.0,24.0)[l]{$H^+$}
\Photon(60.0,35.0)(84.0,35.0){2.0}{4}
\Text(84.0,35.0)[l]{ $\gamma$}
\DashLine(60.0,11.0)(36.0,11.0){1.0}
\Text(36.0,11.0)[r]{$H^0$}
\Photon(60.0,11.0)(84.0,11.0){2.0}{4} 
\Text(84.0,11.0)[l]{ $W^+$}
\end{picture} \ 
{} \qquad\allowbreak
\begin{picture}(120,62)(0,0)
\DashLine(60.0,59.0)(36.0,59.0){1.0}
\Text(36.0,59.0)[r]{$H^0$}
\DashArrowLine(60.0,35.0)(60.0,59.0){1.0} 
\Text(63.0,48.0)[l]{$H^+$}
\Photon(60.0,59.0)(84.0,59.0){2.0}{4} 
\Text(84.0,59.0)[l]{ $W^+$}
\DashArrowLine(60.0,11.0)(60.0,35.0){1.0} 
\Text(63.0,24.0)[l]{$H^+$}
\Photon(60.0,35.0)(84.0,35.0){2.0}{4} 
\Text(84.0,35.0)[l]{ $\gamma$}
\DashLine(60.0,11.0)(36.0,11.0){1.0}
\Text(36.0,11.0)[r]{$H^0$}
\Photon(84.0,11.0)(60.0,11.0){2.0}{4} 
\Text(84.0,11.0)[l]{ $W^-$}
\end{picture} \ 
{} \qquad\allowbreak 
\begin{picture}(120,62)(0,0)
\DashLine(60.0,59.0)(36.0,59.0){1.0}
\Text(36.0,59.0)[r]{$H^0$}
\DashArrowLine(60.0,59.0)(60.0,11.0){1.0} 
\Text(63.0,36.0)[l]{$H^+$}
\Photon(84.0,59.0)(60.0,59.0){2.0}{4} 
\Text(84.0,59.0)[l]{ $W^-$}
\DashLine(60.0,11.0)(36.0,11.0){1.0}
\Text(36.0,11.0)[r]{$H^0$}
\Photon(60.0,11.0)(84.0,35.0){2.0}{4} 
\Text(84.0,35.0)[l]{ $W^+$}
\Photon(60.0,11.0)(84.0,11.0){2.0}{4} 
\Text(84.0,11.0)[l]{ $\gamma$}
\end{picture} \ 
{} \qquad\allowbreak \vspace{20pt}
\begin{picture}(120,62)(0,0)
\DashLine(60.0,35.0)(36.0,35.0){1.0}
\Text(36.0,35.0)[r]{$H^0$}
\DashArrowLine(60.0,35.0)(60.0,11.0){1.0} 
\Text(63.0,24.0)[l]{$H^+$}
\Photon(84.0,59.0)(60.0,35.0){2.0}{4} 
\Text(84.0,59.0)[l]{ $W^-$}
\Photon(60.0,35.0)(84.0,35.0){2.0}{4} 
\Text(84.0,35.0)[l]{ $\gamma$}
\DashLine(60.0,11.0)(36.0,11.0){1.0}
\Text(36.0,11.0)[r]{$H^0$}
\Photon(60.0,11.0)(84.0,11.0){2.0}{4} 
\Text(84.0,11.0)[l]{ $W^+$}
\end{picture} \ 
{} \qquad\allowbreak
\begin{picture}(120,62)(0,0)
\DashLine(60.0,59.0)(36.0,59.0){1.0}
\Text(36.0,59.0)[r]{$H^0$}
\DashArrowLine(60.0,11.0)(60.0,59.0){1.0} 
\Text(63.0,36.0)[l]{$H^+$}
\Photon(60.0,59.0)(84.0,59.0){2.0}{4} 
\Text(84.0,59.0)[l]{ $W^+$}
\DashLine(60.0,11.0)(36.0,11.0){1.0}
\Text(36.0,11.0)[r]{$H^0$}
\Photon(84.0,35.0)(60.0,11.0){2.0}{4} 
\Text(84.0,35.0)[l]{ $W^-$}
\Photon(60.0,11.0)(84.0,11.0){2.0}{4} 
\Text(84.0,11.0)[l]{ $\gamma$}
\end{picture} \ 
{} \qquad\allowbreak
\begin{picture}(120,62)(0,0)
\DashLine(60.0,35.0)(36.0,35.0){1.0}
\Text(36.0,35.0)[r]{$H^0$}
\DashArrowLine(60.0,11.0)(60.0,35.0){1.0} 
\Text(63.0,24.0)[l]{$H^+$}
\Photon(60.0,35.0)(84.0,59.0){2.0}{4} 
\Text(84.0,59.0)[l]{ $W^+$}
\Photon(60.0,35.0)(84.0,35.0){2.0}{4}
\Text(84.0,35.0)[l]{ $\gamma$}
\DashLine(60.0,11.0)(36.0,11.0){1.0}
\Text(36.0,11.0)[r]{$H^0$}
\Photon(84.0,11.0)(60.0,11.0){2.0}{4} 
\Text(84.0,11.0)[l]{ $W^-$}
\end{picture} \ 
{} \qquad\allowbreak \vspace{20pt}
\begin{picture}(120,62)(0,0)
\DashLine(48.0,47.0)(24.0,47.0){1.0}
\Text(24.0,47.0)[r]{$H^0$}
\DashArrowLine(48.0,47.0)(48.0,11.0){1.0} 
\Text(51.0,30.0)[l]{$H^+$}
\Photon(72.0,47.0)(48.0,47.0){2.0}{4} 
\Text(61.0,50.0)[b]{ $W^-$}
\DashLine(48.0,11.0)(24.0,11.0){1.0}
\Text(24.0,11.0)[r]{$H^0$}
\Photon(48.0,11.0)(96.0,11.0){2.0}{8} 
\Text(96.0,11.0)[l]{ $W^+$}
\Photon(96.0,59.0)(72.0,47.0){2.0}{4} 
\Text(96.0,59.0)[l]{ $W^-$}
\Photon(72.0,47.0)(96.0,35.0){2.0}{4} 
\Text(96.0,35.0)[l]{ $\gamma$}
\end{picture} \ 
{} \qquad\allowbreak
\begin{picture}(120,62)(0,0)
\DashLine(48.0,59.0)(24.0,59.0){1.0}
\Text(24.0,59.0)[r]{$H^0$}
\DashArrowLine(48.0,59.0)(48.0,23.0){1.0} 
\Text(51.0,42.0)[l]{$H^+$}
\Photon(96.0,59.0)(48.0,59.0){2.0}{8} 
\Text(96.0,59.0)[l]{ $W^-$}
\DashLine(48.0,23.0)(24.0,23.0){1.0}
\Text(24.0,23.0)[r]{$H^0$}
\Photon(48.0,23.0)(72.0,23.0){2.0}{4} 
\Text(61.0,12.0)[b]{ $W^+$}
\Photon(72.0,23.0)(96.0,35.0){2.0}{4} 
\Text(96.0,35.0)[l]{ $W^+$}
\Photon(72.0,23.0)(96.0,11.0){2.0}{4} 
\Text(96.0,11.0)[l]{ $\gamma$}
\end{picture} \ 
{} \qquad\allowbreak 
\begin{picture}(120,62)(0,0)
\DashLine(48.0,59.0)(24.0,59.0){1.0}
\Text(24.0,59.0)[r]{$H^0$}
\DashArrowLine(48.0,23.0)(48.0,59.0){1.0} 
\Text(51.0,42.0)[l]{$H^+$}
\Photon(48.0,59.0)(96.0,59.0){2.0}{8} 
\Text(96.0,59.0)[l]{ $W^+$}
\DashLine(48.0,23.0)(24.0,23.0){1.0}
\Text(24.0,23.0)[r]{$H^0$}
\Photon(72.0,23.0)(48.0,23.0){2.0}{4} 
\Text(65.0,12.0)[b]{ $W^-$}
\Photon(96.0,35.0)(72.0,23.0){2.0}{4} 
\Text(96.0,35.0)[l]{ $W^-$}
\Photon(72.0,23.0)(96.0,11.0){2.0}{4} 
\Text(96.0,11.0)[l]{ $\gamma$}
\end{picture} \ 
{} \qquad\allowbreak \vspace{20pt}
\begin{picture}(120,62)(0,0)
\DashLine(48.0,47.0)(24.0,47.0){1.0}
\Text(24.0,47.0)[r]{$H^0$}
\DashArrowLine(48.0,11.0)(48.0,47.0){1.0} 
\Text(51.0,30.0)[l]{$H^+$}
\Photon(48.0,47.0)(72.0,47.0){2.0}{4} 
\Text(61.0,50.0)[b]{ $W^+$}
\DashLine(48.0,11.0)(24.0,11.0){1.0}
\Text(24.0,11.0)[r]{$H^0$}
\Photon(96.0,11.0)(48.0,11.0){2.0}{8} 
\Text(96.0,11.0)[l]{ $W^-$}
\Photon(72.0,47.0)(96.0,59.0){2.0}{4} 
\Text(96.0,59.0)[l]{ $W^+$}
\Photon(72.0,47.0)(96.0,35.0){2.0}{4} 
\Text(96.0,35.0)[l]{ $\gamma$}
\end{picture} \ 
{} \qquad\allowbreak 
\begin{picture}(120,62)(0,0)
\DashLine(48.0,23.0)(24.0,35.0){1.0}
\Text(24.0,35.0)[r]{$H^0$}
\DashLine(48.0,23.0)(24.0,11.0){1.0}
\Text(24.0,11.0)[r]{$H^0$}
\Photon(48.0,23.0)(72.0,35.0){2.0}{4} 
\Text(72.0,40.0)[l]{ $W^-$}
\Photon(48.0,23.0)(72.0,23.0){2.0}{4} 
\Text(62.0,12.0)[b]{ $W^+$}
\Photon(72.0,23.0)(96.0,35.0){2.0}{4} 
\Text(96.0,40.0)[l]{ $W^+$}
\Photon(72.0,23.0)(96.0,11.0){2.0}{4} 
\Text(96.0,11.0)[l]{ $\gamma$}
\end{picture} \ 
{} \qquad\allowbreak
\begin{picture}(120,62)(0,0)
\DashLine(48.0,23.0)(24.0,35.0){1.0}
\Text(24.0,35.0)[r]{$H^0$}
\DashLine(48.0,23.0)(24.0,11.0){1.0}
\Text(24.0,11.0)[r]{$H^0$}
\Photon(48.0,23.0)(72.0,35.0){2.0}{4} 
\Text(72.0,40.0)[l]{ $W^+$}
\Photon(48.0,23.0)(72.0,23.0){2.0}{4} 
\Text(62.0,12.0)[b]{ $W^-$}
\Photon(96.0,35.0)(72.0,23.0){3.0}{4} 
\Text(96.0,40.0)[l]{ $W^-$}
\Photon(72.0,23.0)(96.0,11.0){2.0}{4} 
\Text(96.0,11.0)[l]{ $\gamma$}
\end{picture} \ 
{} \qquad\allowbreak \vspace{20pt} 
\begin{picture}(120,62)(0,0)
\DashLine(48.0,47.0)(24.0,59.0){1.0}
\Text(24.0,59.0)[r]{$H^0$}
\DashLine(48.0,47.0)(24.0,35.0){1.0}
\Text(24.0,35.0)[r]{$H^0$}
\DashLine(48.0,47.0)(72.0,47.0){1.0}
\Text(61.0,50.0)[b]{$h$ }
\Photon(96.0,59.0)(72.0,47.0){2.0}{4} 
\Text(96.0,59.0)[l]{ $W^-$}
\Photon(72.0,47.0)(72.0,23.0){2.0}{4} 
\Text(51.0,36.0)[l]{ $W^+$}
\Photon(72.0,23.0)(96.0,35.0){2.0}{4} 
\Text(96.0,36.0)[l]{ $W^+$}
\Photon(72.0,23.0)(96.0,11.0){2.0}{4} 
\Text(96.0,11.0)[l]{ $\gamma$}
\end{picture} \ 
{} \qquad\allowbreak
\begin{picture}(120,62)(0,0)
\DashLine(48.0,47.0)(24.0,59.0){1.0}
\Text(24.0,59.0)[r]{$H^0$}
\DashLine(48.0,47.0)(24.0,35.0){1.0}
\Text(24.0,35.0)[r]{$H^0$}
\DashLine(48.0,47.0)(72.0,47.0){1.0}
\Text(61.0,50.0)[b]{$h$ }
\Photon(72.0,47.0)(96.0,59.0){2.0}{4} 
\Text(96.0,59.0)[l]{ $W^+$}
\Photon(72.0,23.0)(72.0,47.0){2.0}{4} 
\Text(51.0,35.0)[l]{ $W^-$}
\Photon(96.0,35.0)(72.0,23.0){2.0}{4} 
\Text(96.0,35.0)[l]{ $W^-$}
\Photon(72.0,23.0)(96.0,11.0){2.0}{4} 
\Text(96.0,11.0)[l]{ $\gamma$}
\end{picture} \ 
}